\documentclass{aa}

\usepackage{graphicx}
\usepackage{url,fancyvrb,enumitem}
\usepackage{txfonts}

\usepackage[usenames]{xcolor}
\usepackage[pdftex,bookmarks,colorlinks,breaklinks]{hyperref}
\hypersetup{linkcolor=blue,citecolor=blue,filecolor=black,urlcolor=blue}

\usepackage{natbib}
\usepackage{amssymb,emorales-commands}

\begin{document}

  %\title{Towards a hierarchical multi-wavelength catalog of Galactic young stellar objects}
  %\subtitle{I. The potential multiplicity observed with UKIDSS}
  %\titlerunning{Towards a hierarchical multi-wavelength catalog of Galactic young stellar objects. I.}

  \title{Do individual \emph{Spitzer} young stellar object candidates enclose multiple UKIDSS sources?}
  \titlerunning{Do individual \emph{Spitzer} YSO candidates enclose multiple UKIDSS sources?}

  \authorrunning{E. F. E. Morales and T. P. Robitaille}

  \author{Esteban F. E. Morales
          \and
          Thomas P. Robitaille
          }

  \institute{Max-Planck-Institut f{\"u}r Astronomie,
             K\"{o}nigstuhl 17, 69117 Heidelberg, Germany\\
             \email{morales@mpia.de}
             }

  \date{Received / Accepted }

  \abstract
%  {The standard way of combining multi-wavelength observations of young stellar objects (YSOs) detected across available Galactic plane surveys is to associate sources at each wavelength with their nearest neighbors at other wavelengths, missing the high-resolution information of the shortest wavelengths.}
%  {The study of star-formation in the Milky Way has been highly benefited from the development of Galactic plane surveys over the last decade, from the near-infrared to the millimeter. However, the standard way of combining the young stellar objects (YSOs) detected on these data is to associate the nearest sources across the wavelength range, missing the high-resolution information of the shortest wavelengths.}  
  {}
  {We analyze near-infrared UKIDSS observations of a sample of 8325 objects taken from a catalog of intrinsically red sources in the Galactic plane selected in the \emph{Spitzer}-GLIMPSE survey. Given the differences in angular resolution (factor $>2$ better in UKIDSS), our aim is to investigate whether there are multiple UKIDSS sources that might all contribute to the GLIMPSE flux, or there is only one dominant UKIDSS counterpart. We then study possible corrections to estimates of the star formation rate (SFR) based on counts of GLIMPSE young stellar objects (YSOs). This represents an exploratory work towards the construction of a hierarchical YSO catalog.}
  {After performing PSF fitting photometry in the UKIDSS data, we implemented a technique to automatically recognize the dominant UKIDSS sources by evaluating their match with the spectral energy distribution (SED) of the associated GLIMPSE red sources. This is a generic method which could be robustly applied for matching SEDs across gaps at other wavelengths.}
  {We found that most ($87.0 \pm 1.6 \%$) of the candidate YSOs from the GLIMPSE red source catalog have \emph{only one} dominant UKIDSS counterpart which matches the mid-infrared SED (fainter associated UKIDSS sources might still be present). Though at first sight this could seem surprising, given that YSOs are typically in clustered environments, we argue that within the mass range covered by the GLIMPSE YSO candidates (intermediate to high masses), clustering with objects with comparable mass is unlikely at the GLIMPSE resolution. Indeed, by performing simple clustering experiments based on a population synthesis model of Galactic YSOs, we found that although $\sim 60\%$ of the GLIMPSE YSO enclose at least two UKIDSS sources, in general only one dominates the flux.} 
  %This means that no significant corrections are needed to estimates of the SFR of the Milky Way based on the assumption that the GLIMPSE YSOs are individual objects. However, by doing similar experiments using synthetic YSOs to study the presence of unresolved binaries in the GLIMPSE YSO sample (a few of them could be resolved at the UKIDSS resolution), we found that the correction from this effect might be not negligible, and would increase the SFR estimate by a factor $\sim 1.2$--1.3.
  {No significant corrections are needed for estimates of the SFR of the Milky Way based on the assumption that the GLIMPSE YSOs are individual objects. However, we found that unresolved binaries in GLIMPSE objects (a few of them could be resolved at the UKIDSS resolution) have a non-negligible effect, and would increase the SFR estimate by a factor $\sim 1.2$--1.3.}


  \maketitle

\section{Introduction}
\label{sec:introduction}

The new generation of Galactic plane surveys carried out in the last decade, from near-infrared (NIR) to millimeter wavelengths, has started to revolutionize our knowledge of star formation in the Milky Way. In the past, we have been limited to inferring properties of the star formation process from observations of good template regions or a selected sample of individual objects, but these surveys now allow us to explore the whole variety of star-forming environments in the Galactic plane. They are unbiased in spatial coverage within a certain range of coordinates, though still subject to observational limitations such as sensitivity, angular resolution, and (for short wavelengths) interstellar extinction.

Using the mid-infrared (MIR) data from the Galactic Legacy Infrared Mid-Plane Survey Extraordinaire (GLIMPSE), \citet{Robitaille2008} compiled a sample of almost $19\,000$~intrinsically red sources in the inner Galactic plane, thought to be mostly high- and intermediate-mass young stellar objects (YSOs) and asymptotic giant branch (AGB) stars. This population of YSOs was modeled by \citet{RobitailleWhitney2010} to estimate the star formation rate (SFR) of the Milky Way, for the first time using directly individual YSO counts in conjunction with a population synthesis model. On the other hand, recent (sub)mm continuum surveys \citetext{\citealp[Bolocam Galactic Plane Survey,][]{Aguirre2011}; \citealp[ATLASGAL,][]{Schuller2009}} have revealed thousands of potential star-forming cold dust clumps throughout the Galactic plane. By matching the detected (sub)mm sources with samples of ongoing star-formation indicators, namely (massive) YSOs identified in the GLIMPSE or \emph{MSX} MIR surveys, 24~\micron\ point sources from the MIPSGAL survey, \ion{H}{ii} regions, and methanol masers, several studies have characterized the population of these cold clumps, proposing tentative evolutionary stages, identifying ``starless'' clump candidates, and investigating the physical properties that give rise to different modes of star formation in our Galaxy \citep[e.g.,][]{Dunham2011, Tackenberg2012, Urquhart2014}. More recently, the \emph{Herschel} Hi-GAL survey \citep{Molinari2010}, covering wavelengths from 70~\micron\ to 500~\micron, which is the range where the cold dust emission peaks, provides a rich dataset to study the properties of both the embedded YSOs (through the protostellar radiation reprocessed by the surrounding dust) and their envelopes, as well as younger pre-stellar cores \citep[e.g.,][]{Veneziani2013, Elia2013}.

None of these previous studies, however, have considered the potential multiplicity of a single object at a given wavelength when observed at a shorter wavelength providing higher angular resolution. The standard approaches to deal with multi-wavelength data is a plain ``nearest source'' matching to associate sources across different wavelengths or, in the case of sub(mm) clumps, a simple classification of the source based on the type (or absence) of higher-resolution objects that fall within the covered area in the sky, but without explicitly taking into account the multiplicity of these objects. In order to make use of the full high-resolution information of multi-wavelength data, we are working on the development of a hierarchical YSO catalog that will associate multiple sources from a higher-resolution survey with the corresponding (typically fewer) sources in a lower resolution survey. As a long-term plan, our idea is to apply this formalism to the whole set of Galactic plane surveys, from the NIR to the millimeter.

In this paper, we present an exploratory study which constitutes the first step towards the construction of the hierarchical YSO catalog. Instead of constructing a catalog from scratch and using all the Galactic surveys available, we make here two simplifications: first, we only use data from the GLIMPSE survey and the NIR higher-resolution United Kingdom Infrared Deep Sky Survey (UKIDSS); and second, we start from the GLIMPSE individual objects already selected by \citet{Robitaille2008} and check whether they split into multiple sources when seen in UKIDSS. In particular, we investigate for each GLIMPSE object whether there are multiple UKIDSS sources that might all contribute to the GLIMPSE flux, or there is only one dominant UKIDSS counterpart. Though this approach is more similar to the previous studies mentioned above in the sense that it is based on a sample of known objects selected from a given survey to investigate how they look at a higher-resolution survey, to our knowledge the present work is the first study which tries to associate single GLIMPSE objects with multiple UKIDSS sources -- instead of just associating the nearest source -- in a large fraction of the Galactic plane. We have chosen the \citet{Robitaille2008} census as our starting point not only for simplicity and good coverage of the Galactic plane, but also because this sample is highly reliable. As a direct scientific application of this work, we can study the validity of the physical properties derived when assuming that the GLIMPSE YSOs in the Galactic plane are single objects. In particular, in this paper we investigate how clustering and unresolved binaries could affect the \citet{Robitaille2008} sample, and therefore the SFR estimate of \citet{RobitailleWhitney2010}. 

In Section~\ref{sec:observations}, we briefly describe the GLIMPSE observations and the corresponding YSO catalog, as well as the data from the UKIDSS Galactic Plane Survey we used in this work. Section~\ref{sec:photometry} gives details about the PSF fitting photometry we performed on the UKIDSS data, and the quality control of this photometry. In Section~\ref{sec:results} we report the main results of this study, in particular the implementation of a method to automatically identify dominant UKIDSS sources in the spectral energy distribution, and the statistics for the GLIMPSE YSO sample regarding these UKIDSS sources; Appendices~\ref{sec:decision-rules} and \ref{sec:convex} describe more specific aspects of our method, while in Appendices~\ref{sec:statistics-details} and \ref{sec:extended-sample} we give, respectively, further technical details about our statistics and extended results for special cases. In Section~\ref{sec:discussion}, we discuss the nature of possibly multiple UKIDSS sources, the sensitivity of our method to flux changes, and the presence of variable sources; we also present simple simulated YSO populations models \citep[based on the work by][]{RobitailleWhitney2010} to assess the importance of clustering and unresolved binaries in the GLIMPSE YSO sample and compare with the properties of the UKIDSS observations analyzed here, as well as to discuss the implications for SFR estimates. Finally, Section~\ref{sec:conclusions} summarizes the main conclusions of this paper.

\section{Data sets}
\label{sec:observations}

\defcitealias{Robitaille2008}{R08}

\subsection{GLIMPSE YSO candidates}
\label{sec:glimpse-data}

GLIMPSE \citep{Benjamin2003,Churchwell2009} consists of a set of various MIR surveys of the Galactic plane carried out with the InfraRed Array Camera \citep[IRAC,][]{Fazio2004}, at 3.6, 4.5, 5.6, and 8.0~\micron\ and with an angular resolution (FWHM) of $\sim 2''$, on board of the \emph{Spitzer Space Telescope} \citep{Werner2004}. Here we use the census of intrinsically red MIR sources from \citet[hereafter R08]{Robitaille2008} which is specifically based on the GLIMPSE~I and II surveys. These two surveys cover the $(\ell,b)$ ranges: $5\degr < |\ell| \le 65\degr$ and $|b| \le 1\degr$; $2\degr < |\ell| \le 5\degr$ and $|b| \le 1.5\degr$; $|\ell| \le 2\degr$ and $|b| \le 2\degr$, comprising a total of 274 square degrees.

The GLIMPSE~I and II point source catalogs were used by \citetalias{Robitaille2008} to compile a highly reliable census of $18\,949$ sources selected by their MIR red color, namely $[4.5] - [8.0] \ge 1$, and additional brightness criteria that were carefully chosen to minimize the effects of position-dependent sensitivity and saturation: $13.89 \ge [4.5] \ge 6.50$ and $9.52 \ge [8.0] \ge 4.01$. \citetalias{Robitaille2008} further filtered their selection by imposing a set of quality criteria (their Equation (4)) to reduce the contamination by bad photometry and spurious detections. The \citetalias{Robitaille2008} catalog provides improved coordinates and photometry at 4.5 and 8.0~\micron\ (each source was visually examined to ensure that the improved photometry could be trusted), and magnitudes at 24~\micron\ obtained from photometry performed on images of the MIPSGAL survey \citep{Carey2009}, together with the original GLIMPSE photometry at 3.6 and 5.8~\micron, and $JHK_{\rm s}$ magnitudes from the 2MASS survey \citep[included in the GLIMPSE point source catalog]{Skrutskie2006}.

The sources from the \citetalias{Robitaille2008} sample consist mostly of YSOs (with an estimated fraction 50\%--70\%) and AGB stars (estimated fraction of 30\%--50\%). The separation of the sources into YSOs and AGB stars was based on simple color and magnitude selection criteria \citepalias[Equations (7)-(9) by][]{Robitaille2008}, and was only approximate. However, supported by the angular distribution of both samples in the Galactic plane (uniform for AGB stars, and more clustered for YSOs), \citetalias{Robitaille2008} claimed that this separation should be sufficient to estimate the relative fraction of AGB stars and YSOs in their sample.

One aspect of the GLIMPSE point source catalog and the photometry performed by \citetalias{Robitaille2008}, which is relevant for the analysis presented here, is the processing of close and/or blended sources. Although the GLIMPSE pipeline is based on the point spread function (PSF) fitting photometry program \daophot, which in principle is able to extract fluxes of overlapping sources (see Section~\ref{sec:photometry}), the fact that the IRAC PSFs are undersampled (the camera has 1.2\arcsec\ pixels) makes the splitting of the flux between close sources very inaccurate. Therefore, all the detections within 2\arcsec\ in an individual IRAC frame were ``lumped'' into one source before the in-band merge (merging sources in the same band from different frames) and the cross-band merge (B.~Babler 2016, private communication; see also the GLIMPSE~I~v2.0 data release document\footnote{\url{http://irsa.ipac.caltech.edu/data/SPITZER/GLIMPSE/doc/glimpse1_dataprod_v2.0.pdf}}). Due to the in-band merge and the migration of some positions during the cross-band merge, a few cases of sources closer than 2\arcsec\ are still present in the more complete GLIMPSE point source ``archive'' (these cases are filtered out in the ``catalog''), but in general objects within 2\arcsec\ are not separated and are considered a unique source by the GLIMPSE photometry. \citetalias{Robitaille2008} did remove from their sample some cases of blended sources which produced unreliable measurements during their manual photometry; however, we noticed by visual inspection that most of those blended sources are separated by more than 2\arcsec. In this paper, we thus assume that multiple objects within 2\arcsec\ are in general included and identified as only one source in the \citetalias{Robitaille2008} sample.

Related to what discussed above, one of the quality criteria imposed by \citetalias{Robitaille2008} on the GLIMPSE catalog to select their sample was a close source flag equal to zero, which means that each source has to be free of neighbors between 2\arcsec and 3\arcsec from its position. Although this criterion could have removed some YSOs which were in very dense environments, we note that if we eliminate this restriction the number of selected sources would only increase by $\sim 10\%$ with respect to the number of sources in the \citetalias{Robitaille2008} sample, so that the effect of this particular selection is not significant.

\subsection{The UKIDSS Galactic Plane Survey}
\label{sec:ukidss-data}

In this work, we use data from UKIDSS \citep{Lawrence2007}, in particular the UKIDSS Galactic Plane Survey \citep[GPS,][]{Lucas2008}. This survey was conducted with the Wide Field Camera \citep[WFCAM,][]{Casali2007} mounted on the United Kingdom Infrared Telescope (UKIRT), on Mauna Kea, Hawaii, providing images in the $J$ (1.25~\micron), $H$ (1.63~\micron), and $K$ (2.20~\micron) filters, with an angular resolution (FWHM) of $\sim 0.9''$. Here we use the UKIDSS GPS products from the Data Release~8 (DR8), which covers 975.5 square degrees (in all filters) of the northern and equatorial Galactic plane, including most of the area covered by GLIMPSE~I and II surveys for $\ell \ge -2$. DR8 was the latest data release during the time that the work presented here was carried out. The GPS data from DR10 were released on January, 2015, and will be used on future papers, together with data from the VISTA Variables in the V\'{i}a L\'actea survey \citep[VVV,][]{Minniti2010}, which covers the southern Galactic plane.

\begin{figure*}[!t]
\centering
\includegraphics[width=0.99\textwidth]{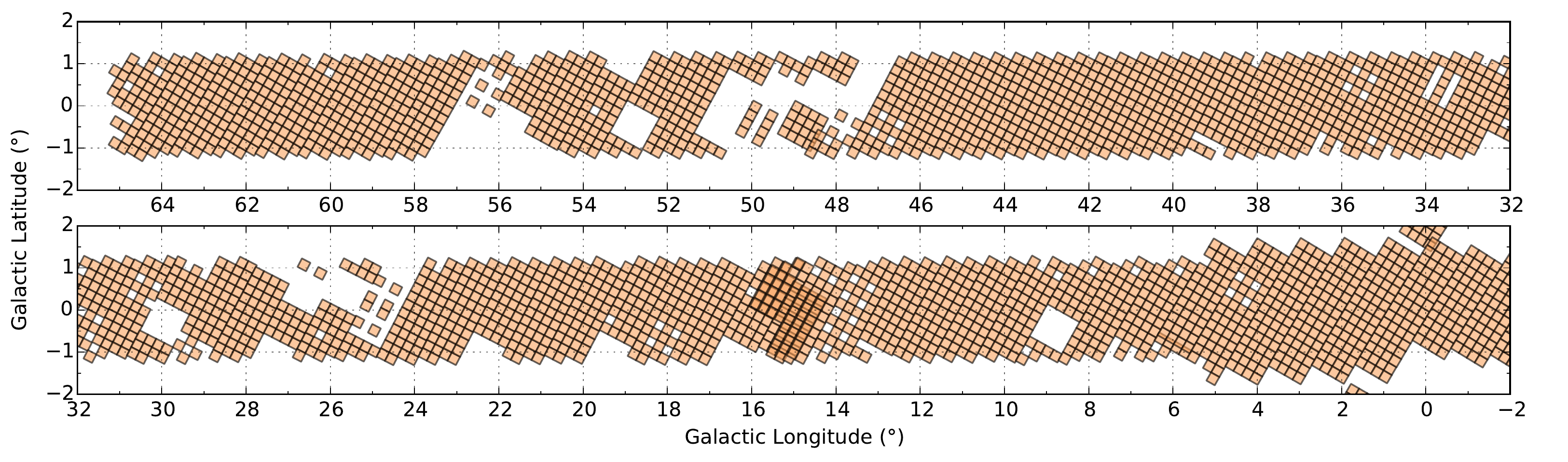}
\caption{Positions of UKIDSS GPS DR8 frame sets covering the GLIMPSE I/II observed area.}
\label{fig:dr8-coverage}
\end{figure*}

The UKIDSS observations are reduced and calibrated at the Cambridge Astronomical Survey Unit (CASU)\footnote{\url{http://casu.ast.cam.ac.uk/}}, including the generation of a point source catalog for each individual co-added ``Multiframe'' (see below). These data are then organized in the WFCAM Science Archive into a SQL database \citep[WSA,][]{Hambly2008}\footnote{\url{http://surveys.roe.ac.uk/wsa/index.html}}, where the catalog detections in the different filters are merged. A Multiframe is the fundamental unit of the UKIDSS database, and consists of four $\sim 13.65\arcmin \times 13.65\arcmin$ spatially separated patches of the sky observed in one filter, inherited from the four array detectors of WFCAM. Each patch is called ``MultiframeDetector'' in the WSA nomenclature. In this paper, we directly use the final co-added images known as \emph{leavstack} Multiframes \citep[see Appendix A1 in][for more details]{Lucas2008}, but for a given position we access them through the identification numbers assigned by the band-merging process in the WSA, which groups individual well-aligned frames observed in different filters into a \emph{frame set}, i.e., a set of MultiframeDetectors \citep[see][]{Hambly2008}. By doing this, we ensure the access to the best set of UKIDSS images for a given position, and in cases of bright diffuse background emission, to the images that were already processed with a nebulosity-filtering algorithm \citep[originally designed to improve the extracted catalogs,][]{Irwin2010}.

Figure~\ref{fig:dr8-coverage} shows the positions and coverage of all UKIDSS GPS DR8 frame sets overlapping the GLIMPSE~I/II area, after excluding a few frames with poor astrometry\footnote{For details on this issue, check \url{http://surveys.roe.ac.uk/wsa/gpsAstrometryDR8.html}}. We downloaded all the Multiframe images associated with these frame sets, and extracted a point source catalog per frame set by automatically querying the on-line UKIDSS catalog available at the WSA (these catalogs were only used for PSF stars selection, as explained in Section~\ref{sec:PSF}). We then searched for the best frame set available for each object from the \citetalias{Robitaille2008} sample and stored the associated identification number (\verb|frameSetID|). Again, this was done by automatically querying the WSA UKIDSS catalog at each position within a $12\arcsec$ radius, and by selecting the \verb|frameSetID| associated with the nearest of the sources satisfying some quality criteria \citep[following Appendix A3 by][]{Lucas2008}, namely: select only primary detections for duplicates in overlapping frames, exclude detections classified as noise, and select only sources without serious quality issues ($\verb|ppErrBits| < 256$ in all filters).

In total, $8\,325$ objects from the \citetalias{Robitaille2008} sample, which represent 86\% of all \citetalias{Robitaille2008} objects with $\ell \ge -2$, have available UKIDSS GPS DR8 images in the three $JHK$ filters.

\section{UKIDSS photometry}
\label{sec:photometry}

\begin{figure*}[!t]
\centering
\includegraphics[width=0.99\textwidth]{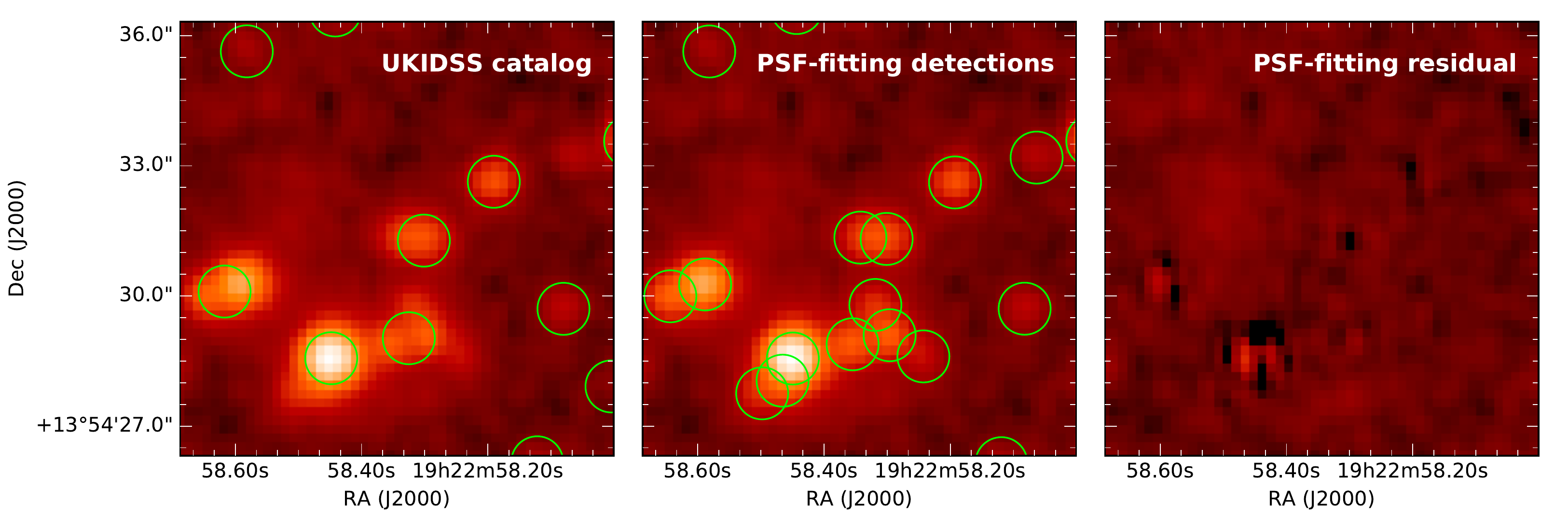}
\caption{Comparison between the point source catalog generated by the WFCAM pipeline (based on aperture photometry) and the sources detected and measured in this work by performing PSF fitting photometry, for the \citetalias{Robitaille2008} object SSTGLMC~G048.8731$-$00.5091. \emph{Left:} UKIDSS GPS $H$-band image, overlaid with the positions of the point sources from the UKIDSS catalog. \emph{Middle:} The same image overlaid with the positions of the objects detected by our PSF fitting photometry. \emph{Right:} Residual image of the PSF fitting photometry, i.e., after subtraction of scaled PSFs at the positions of the detected sources.}
\label{fig:aperture-psf-phot}
\end{figure*}

The UKIDSS point source catalog produced by the WFCAM pipeline is based on aperture photometry only. \citet{Lucas2008} claim that overlapping apertures are properly treated in the pipeline, and that no significant differences in precision are found in test results comparing this type of aperture photometry with PSF fitting photometry. However, we do expect the latter to give an improved accuracy and completeness in the presence of blended stars or in very crowded regions, as has been already shown in previous studies using UKIDSS data of star clusters \citep[e.g.,][]{SteadHoare2011,Longmore2011}. Since we are interested in the possible multiplicity of YSOs seen as individual sources in GLIMPSE, we performed PSF fitting photometry on the UKIDSS images, in order to detect and properly separate the emission of multiple overlapping sources at the position of each \citetalias{Robitaille2008} object. As an example, this higher completeness can be clearly noted in Figure~\ref{fig:aperture-psf-phot}, which compares the detections found in the UKIDSS point source catalog with the ones in our PSF fitting photometry for a small field centered on one \citetalias{Robitaille2008} source.

We used the standard PSF fitting routines of the \daophot\ package \citep[including the programs \allstar\ and \allframe;][]{Stetson1987, Stetson1994}, which were run in a completely automatic way through a custom Python wrapper. Here, we briefly describe some points particularly related to the UKIDSS photometry we performed in this work; for more details regarding the general functioning of \daophot, we refer the reader to the original papers or the \daophot\ user's manual\footnote{It can be found at \url{http://www.astro.wisc.edu/sirtf/daophot2.pdf}}. It should be noted that before running these routines, the UKIDSS images were processed with the \emph{dribbling} method, i.e., convolved with a simple matrix in order to redistribute the original 0.4\arcsec\ pixel flux into the new 0.2\arcsec\ pixel grid after the \emph{interleaving} process\footnote{See \url{http://apm49.ast.cam.ac.uk/surveys-projects/wfcam/technical/interleaving} for details.}.

\subsection{Construction of the PSFs}
\label{sec:PSF}

We constructed a spatially-varying PSF for each individual detector of every Multiframe in our data set, using a selection of bright stars from the corresponding point source catalog. The selection was based on the quality criteria applied to generate the ``most reliable, low completeness'' sample in Appendix A3 of \citet{Lucas2008}, together with the stricter condition $\verb|ppErrBits| \le 16$ in all filters (only the \emph{deblended} error flag is allowed). We generated slightly different lists of PSF stars per filter by imposing individual magnitude cuts $J > 13.25$, $H > 12.75$ and $K > 12$, which are the conservative saturation limits recommended by \citet{Lucas2008}. The PSF stars were also required to be free of brighter or saturated neighbors within a radius equal to \textsc{fitting radius} + \textsc{psf radius} (parameters in \daophot; we use values of 4 and 20 pixels, respectively), in order to emulate the selection of PSF stars done by the routine \textsc{pick} of \daophot.

From this restricted very high-quality sample, we selected the 100 brightest stars (in each frame and each filter) to construct the PSF, which was allowed to vary quadratically across the field\footnote{In a very few cases (19 out of 2125 frame sets used, representing fields with severe source crowding close to the Galactic center), there are less than 100 stars satisfying all the quality criteria in the $K$ filter; however, only two frame sets have less than 50 selected stars in $K$ (namely, 29 and 36 stars), for which we construct a linearly varying PSF.}. The PSF was derived iteratively following a standard sequence of steps: construct a preliminary PSF, do PSF fitting photometry for all neighbor sources of the PSF stars, subtract the neighbors from the image using properly scaled PSFs, and derive an improved PSF from this clean image. Our script ran for 5 iterations, and the complexity of the PSF was gradually increased from a purely analytic constant function to an analytic function plus quadratically varying empirical corrections.

We measured the size (FWHM) of the PSF evaluated at 9 uniformly distributed positions in every individual detector. The sizes are very similar for all the UKIDSS bands, with a mean of $\sim 0.9\arcsec$ and a dispersion of $\sim 0.1\arcsec$. For 96\% of all the measurements in our data set, the sizes are within the range $[0.7\arcsec$,$1.1\arcsec]$.

% MergedClass == -1 is probably satisfied by all sources with pStar > 0.99

\subsection{PSF fitting photometry}
\label{sec:daophot}

We then proceeded to generate a list of point sources with measured $JHK$ magnitudes for every \citetalias{Robitaille2008} object, by carrying out PSF photometry in a small $30\arcsec \times 30\arcsec$ field centered on the object position. The UKIDSS images in the different bands were accessed through the previously stored \verb|frameSetID|. Because we needed the full extent of the frame in order to properly apply the quadratic variations of the PSF, pixels outside the selected $30\arcsec \times 30\arcsec$ field were simply masked, rather than cut out from the image. For every filter, a preliminary catalog was produced by running 3 iterations of a sequence of the \daophot\ routines \textsc{find}, \textsc{photometry}, and \allstar. In the first iteration, we used \textsc{find} to search the image for star-like objects exceeding the local background by at least $4\sigma$, where $\sigma$ is the noise of the image (see below); rough aperture magnitudes for all the detected objects were estimated with \textsc{photometry}; and we then calculated more precise positions and magnitudes through the simultaneous multi-PSF fitting performed by \allstar. In the subsequent iterations, we ran \textsc{find} in the residual image produced by \allstar\ (this time with a threshold of $5\sigma$ to account for the increased noise after the subtraction) in order to identify previously undetected stars that were blended with brighter companions; we estimated aperture magnitudes of the newly detected objects and appended the new star list to the output list of the previous \allstar\ run; and we repeated the PSF fitting photometry of \allstar\ with this extended list as input.

This typical usage of the \daophot\ routines was slightly modified in our pipeline by the incorporation of two custom adjustments. First, the total noise originally used in \daophot, estimated as Poisson noise plus readout noise, could be significantly underestimated in very crowded regions, where the source confusion noise is important (we verified that this was indeed the case in fields close to the Galactic center). Thus, we empirically derived the noise $\sigma$ and the mode (sky level) of the image by a two-step algorithm: iterative $\sigma$-clipping for preliminary estimates, and iterative calculation of the mode through a kernel density estimator, and of the standard deviation by reflecting the pixel values below the mode. We then conservatively used the greatest value of $\sigma$ among the theoretical and empirical estimates.

The second modification in the \textsc{find}/\allstar\ cycle was filtering spurious sources after \emph{every} \allstar\ run, which are unavoidably detected in some cases due to very localized extended emission (which, due to its local nature, was not completely corrected by the nebulosity filter of the WFCAM pipeline) and/or imperfect PSF subtraction from the previous iteration. There are two quality parameters given as part of the \allstar\ output, \emph{sharp} (hereafter, represented by $r_0$) and $\chi$, which represent, respectively, a first-order estimate of the intrinsic radius of the object (in pixels), and a dimensionless measure of the goodness of the PSF fitting. More details on this can be found in the \daophot\ user's manual and in \citet{Stetson2003}. After doing visual checks in some critical fields, we found that a good filtering is achieved, without compromising the detection of blended objects in the new iterations, if we apply thresholds on $r_0$ and on the flux error $\Delta F$ relative to the flux $F$. We used the criteria: $|r_0| \le 1.5$ and $\Delta F/F \le 50\%$ for all detections, and $\Delta F/F \le 15\%$ for the new detections found in the current iteration. We did not impose any limit in the other quality index $\chi$, because some sources can have a high $\chi$ due to the initially bad residual produced by the non-detection of nearby blended sources that will be detected in the subsequent iterations.

After deriving an individual list of point sources for each UKIDSS filter, a preliminary band-merged list was obtained by cross matching the detections with \textsc{daomaster}. At the same time, this routine provides geometrical transformations relating the coordinate systems of the different frames, though in this particular data set the corresponding frames are reasonably well aligned and therefore these transformations represent slight corrections only. We calculated initial coordinate transformations as input for \textsc{daomaster} using the World Coordinate System (WCS) information of the image headers. A final band-merged list of detected UKIDSS point sources for every \citetalias{Robitaille2008} object was produced by running the \allframe\ program, which performs multi-PSF fitting photometry on the three $JHK$ images simultaneously. We converted the data count fluxes to calibrated magnitudes using Equation (C1) of \citet{Hodgkin2009} and the calibration parameters from the image headers.

\subsection{Quality filtering and flags}
\label{sec:quality}

Although we are only interested in the UKIDSS sources in the close vicinity of every \citetalias{Robitaille2008} object, the PSF fitting photometry was performed in larger fields ($30\arcsec \times 30\arcsec$) in order to properly cross-match a significant number of detections in the different bands and derive more precise coordinate transformations between the frames. We can use the final coordinates of the sources measured by \allframe\ in the different filters and the WCS information in the headers to test the relative precision of the UKIDSS astrometry. We found relative differences, between the positions of the same UKIDSS sources in the different bands, larger than 1 pixel (0.2\arcsec) for only 116 \citetalias{Robitaille2008} objects, which were discarded for further analysis.

Hereafter, we will consider all the detected UKIDSS sources within a radius of 2\arcsec\ from each \citetalias{Robitaille2008} object, so that we are only taking into account sources that might contribute to the total flux detected in GLIMPSE (see Section~\ref{sec:glimpse-data}). First, we checked for the presence of saturation in the images possibly affecting these sources. Saturated pixels were defined as having $> \min\{35\,000, \verb|SATURATE|\}$ counts, where \verb|SATURATE| is the average saturation level provided in the image header, and $35\,000$ is a conservative reference value given by \citet{Lucas2008}. Objects from the \citetalias{Robitaille2008} sample that have saturated pixels in at least one of the three UKIDSS $JHK$ images within the 2\arcsec-radius circle were removed from the final sample. On the other hand, cases of saturation occurring outside that circle but within one PSF radius from the perimeter were simply flagged as objects with ``peripheral saturation'', in which the UKIDSS photometry of the potentially associated sources could have been affected by the subtraction of poorly fitted PSF wings of nearby saturated stars. We discarded these objects from the sample, except for very specific parts of the analysis in which we explicitly mention the inclusion of them. We found a total of 2218 \citetalias{Robitaille2008} objects saturated in UKIDSS within 2\arcsec, 2012 of which (91\%) consistently have 2MASS magnitudes brighter than at least one of the conservative UKIDSS saturation limits recommended by \citet{Lucas2008}\footnote{Most of the remaining 9\% of the objects have no 2MASS magnitudes available, so either they were undetected in 2MASS (because of variability or due to the much lower sensitivity of 2MASS with respect to UKIDSS, especially towards the Galactic center), or they have unreliable 2MASS fluxes which were then rejected by \citetalias{Robitaille2008}.}: $J < 13.25$, $H < 12.75$ or $K_{\rm s} < 12$. The useful sample is then reduced to 5991 targets, which are divided into 636 objects with peripheral saturation and 5355 objects not affected by saturation at all.

\begin{figure*}[!t]
\centering
\includegraphics[width=0.99\textwidth]{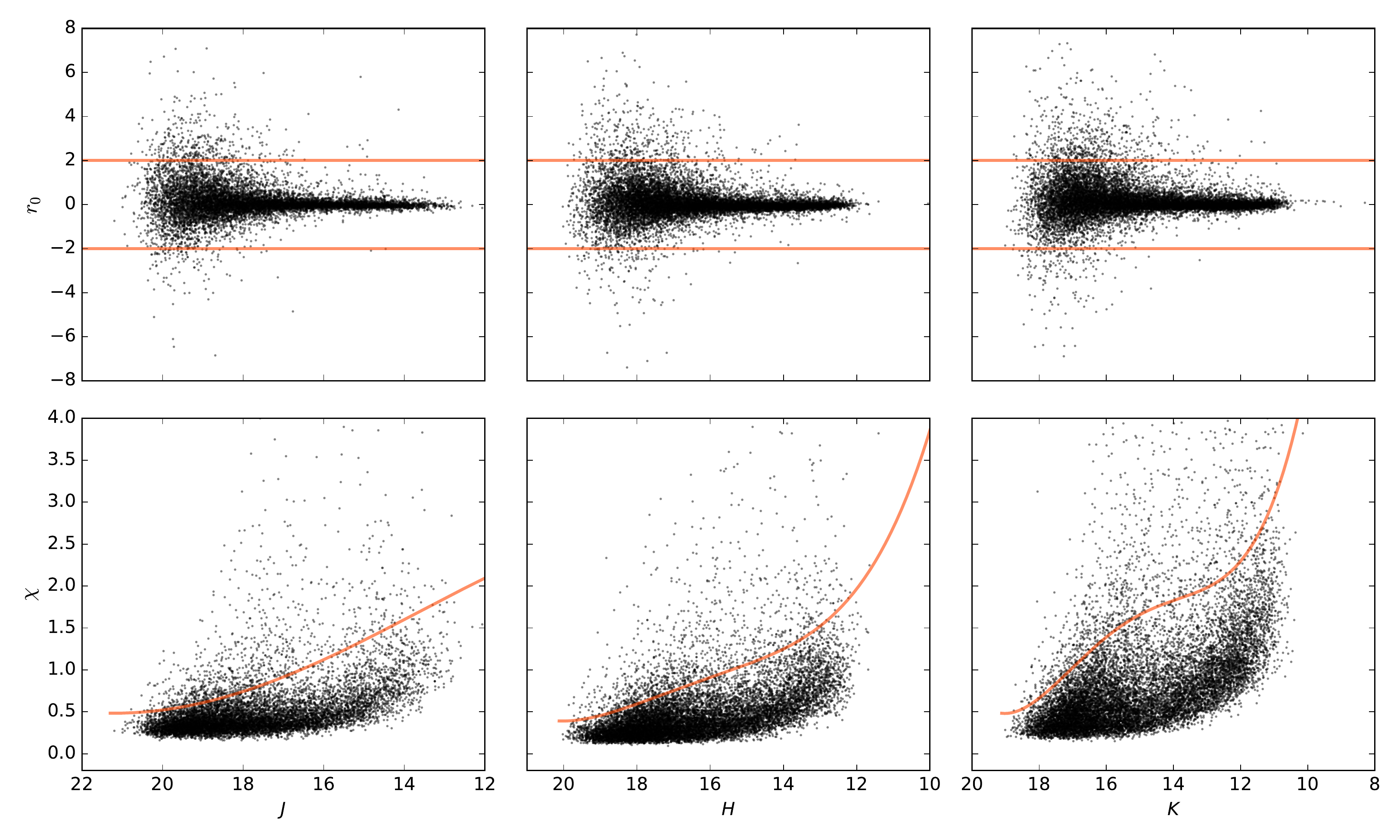}
\caption{\daophot\ quality indices $r_0$ and $\chi$ as a function of the magnitude, for all sources detected in each UKIDSS band, within 2\arcsec\ of non-saturated \citetalias{Robitaille2008} objects. The different lines show the thresholds defining the bad quality flag in a particular filter: $|r_0| = 2$ in the top panels, and the curve $\chi_{\rm lim}(m)$ in the bottom panels (derived empirically as described in the text). A source was considered as having bad quality photometry in given band if $|r_0| > 2$ or $\chi > \chi_{\rm lim}(m)$.}
\label{fig:quality}
\end{figure*}

For the $14\,291$ detected UKIDSS sources within 2\arcsec\ of these 5991 \citetalias{Robitaille2008} objects, we computed individual $3\sigma$-detection limits. Output fluxes below these limits can exist in \allframe\ due to the simultaneous PSF fitting in all bands, but in this work we used instead the computed detection limits (i.e., lower limit magnitudes). For a given source, we converted a $3\sigma$-peak in counts to a calibrated magnitude by using the PSF at that position and the same calibration process applied to the source. The UKIDSS sensitivity varies considerably in our data set due to the highly irregular confusion level throughout the inner Galactic plane \citep[see also Section 3.1 of][]{Lucas2008}, reaching in some uncrowded regions $3\sigma$-detections limits of $\sim 22$, 21, and 20, respectively, in the $K$, $H$, and $K$ bands. However, conservative upper limit sensitivities (i.e., lower limit $3\sigma$-magnitudes) can be set at $J \simeq 19$, $H \simeq 18$, and $J \simeq 17$, for most of the sources in our sample.

We also defined a bad quality flag for this whole sample of UKIDSS sources, using the \daophot\ quality indices $r_0$ and $\chi$ introduced in Section~\ref{sec:daophot}, which are also part of the \allframe\ output for the different bands. Note that this is an independent step to the quality filtering already carried out after every \allstar\ run, which was intended to generate a reliable star list as input for \allframe\ (the final photometry and quality indices were then recalculated by \allframe). In Figure~\ref{fig:quality}, we plot these quality parameters as a function of the magnitude\footnote{The values of $\chi$ do not converge to unity for faint magnitudes as expected, but closer to $\sim 3/8$, representing approximately the noise reduction after the \emph{dribbling} process (see Section~\ref{sec:photometry}). This occurs because $\chi$ is scaled with respect to the theoretical noise estimated by \daophot, which is equivalent (for regions without source confusion) to the noise of the images with the original 0.4\arcsec\ pixel flux.} for all sources with magnitudes brighter than the individual 3$\sigma$-detection limits previously computed, corresponding to a total of $10\,148$, $13\,061$, and $13\,835$ sources, respectively, in the $J$, $H$, and $K$ bands. A source was considered as having bad quality photometry in a particular filter if $|r_0| > 2$ or $\chi > \chi_{\rm lim}(m)$, where $\chi_{\rm lim}(m)$ is an empirically derived threshold that depends on the magnitude $m$, following \citet{Stetson2003}. We note that all the remaining sources satisfy $\Delta F/F \le 50\%$, so that it was not necessary to explicitly impose this condition in this particular sample. We slightly increased the tolerance in $|r_0|$ with respect to the \allstar\ quality filtering, because this time we were only defining flags for sources already validated as non-spurious. To derive the threshold $\chi_{\rm lim}(m)$, we did not use an analytical expression adjusted by eye on the $m$-$\chi$ plot as in \citet{Stetson2003}, given that our plot does not exhibit a trivial dependency, especially in the $K$ band. We instead computed $\chi_{\rm lim}(m)$ as the limit below which we retain the best 90\% of the sources in every filter. Specifically, we implemented the following steps: construction of a surface density $m$-$\chi$ plot using a kernel density estimator; for a grid of 100 values of $m$, calculation of the value $\chi_{0.9}$ at which the cumulative distribution function is 0.9; and finally fit of a 4-degree polynomial to the resulting $(m,\chi_{0.9})$ pairs. The bad quality thresholds in $r_0$ and $\chi$ are also shown in Figure~\ref{fig:quality} for each band.

Applying the above criteria, we flagged 1296, 1537, and 1666 UKIDSS sources, respectively, in the $J$, $H$, and $K$ bands. These typically correspond to different sources in each filter, and only 321 sources were flagged in all three bands.

\section{Results}
\label{sec:results}

\subsection{SED exploration}
\label{sec:sed-exploration}

Detecting the UKIDSS sources within 2\arcsec\ from each of the \citetalias{Robitaille2008} objects is not enough to address the possible multiplicity of these objects, since some of the identified sources could be unrelated foreground/background stars along the same line of sight. Ideally, one would have to determine somehow if these UKIDSS sources are YSO candidates by themselves. Because YSOs are harder to distinguish from field stars via NIR colors only \citep[see, e.g., Section 4.1 of][]{Lucas2008}, an optimal method would be the execution of PSF fitting photometry on the GLIMPSE images using the positions of the UKIDSS sources as input. In this way, we would obtain the contribution of these sources to the flux at the different GLIMPSE wavelengths, so that we would have access to their MIR colors (or colors combining NIR and MIR bands). This is beyond the scope of this particular paper, but analogous approaches will be addressed in the context of the hierarchical YSO catalog we are developing in the future (see Section~\ref{sec:introduction}).

\begin{figure*}[!t]
\centering
\includegraphics[width=0.99\textwidth]{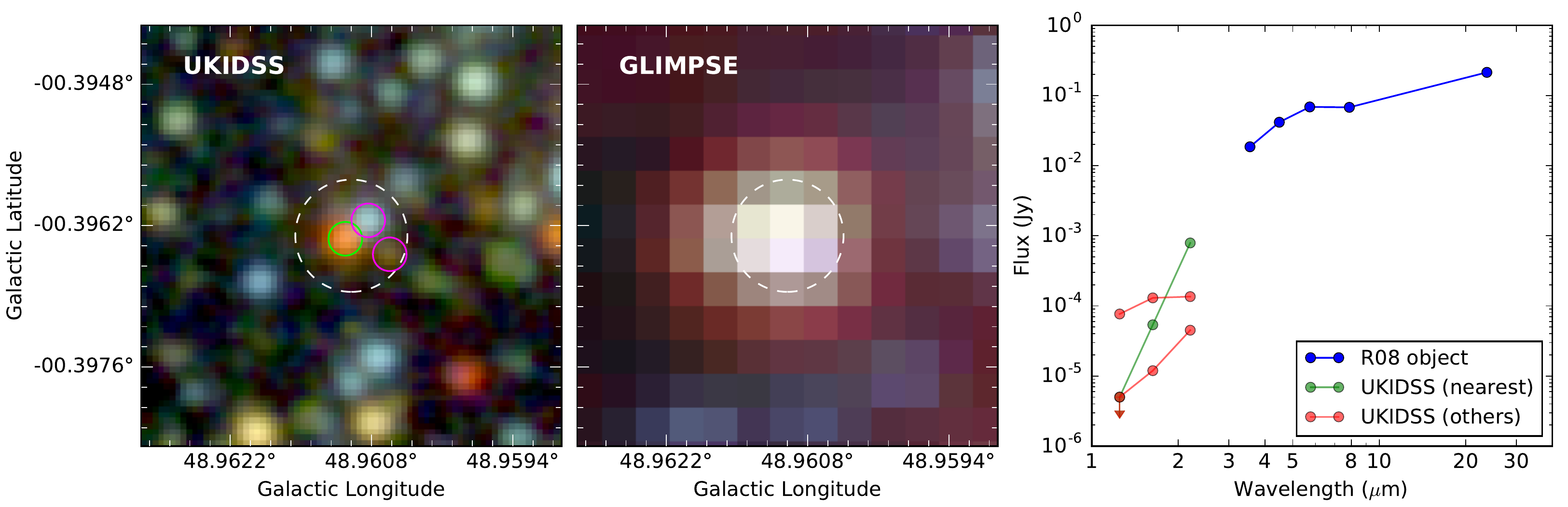}
\caption{Illustration of the usefulness of the SEDs comparison to identify dominant UKIDSS sources. This example shows the \citetalias{Robitaille2008} object SSTGLMC~G048.9610$-$00.3963 and the three detected UKIDSS sources within a radius of 2\arcsec, which is indicated by the dashed-line circle. \emph{Left:} UKIDSS $JHK$ three-color image overlaid with the positions of the UKIDSS sources as open circles (green for the nearest in angular separation, and magenta for the others). \emph{Middle:} GLIMPSE 3.6, 4.5, 8.0~\micron\ three-color image of the same field. \emph{Right:} SED of the \citetalias{Robitaille2008} object at 3.6, 4.5, 5.8, 8.0, and 24~\micron\ wavelengths (blue points), plotted together with the SEDs of the three UKIDSS sources in the $J$, $H$ and $K$ bands (green points for the nearest source, and red points for the others).}
\label{fig:sed-usefulness}
\end{figure*}

In this work, we focus on a simpler -- but also useful -- approach, consisting in the identification of only such UKIDSS sources that have a significant or \emph{dominant} contribution to the observed flux in the GLIMPSE bands. We noticed that comparing the NIR SEDs of the UKIDSS sources with the MIR SED of the corresponding \citetalias{Robitaille2008} object was remarkably helpful to find the dominant sources. As an example, Figure~\ref{fig:sed-usefulness} shows UKIDSS and GLIMPSE three-color images of one particular \citetalias{Robitaille2008} object, together with its MIR SED (3.6--24 \micron) and the NIR SEDs ($JHK$) of the three detected UKIDSS sources. By examining the UKIDSS image and the SEDs comparison plot, it is clear that the nearest (in projected angular distance) UKIDSS source matches nicely the MIR SED and is probably the main contributor to the GLIMPSE fluxes, whereas the others are a much fainter reddened source that does not match well the MIR SED, and an evident foreground star with a flat NIR slope. In Section~\ref{sec:sed-match}, we will describe the methodology we implemented to automatically recognize these UKIDSS sources having a ``good match'' with the MIR SED and therefore being the dominant counterparts of the objects from the \citetalias{Robitaille2008} sample.

\subsection{Quantifying the SED match}
\label{sec:sed-match}

By inspecting several SED comparison plots similar to the one shown in the right panel of Figure~\ref{fig:sed-usefulness}, we realized that the UKIDSS dominant counterparts are typically characterized by a smooth transition between their NIR SED and the MIR SED of the corresponding \citetalias{Robitaille2008} objects, which does not occur at all for much fainter reddened sources or foreground/background stars. To evaluate whether or not we can trace a smooth curve crossing the NIR and MIR points of the combined SED, particularly through the NIR-MIR interface, we computed the cubic spline representation of the SED constructed for each UKIDSS source (and the associated \citetalias{Robitaille2008} object). We then quantified how similar was this spline to a simple quadratic function fitted over the 4 ``middle points'' of the SED defining the NIR-MIR transition (typically the $H$, $K$, 3.6~\micron, and 4.5~\micron\ filters), by computing the mean ratio of both curves. Given the high dynamic range of the SEDs, all calculations were done in $(\log_{10}(\lambda), \log_{10}(F_\nu))$ space, where $\lambda$ is the wavelength and $F_\nu$ is the flux density; we adopted the zero-point fluxes given by \citet{Hewett2006} for the UKIDSS band, and by \citetalias{Robitaille2008} for the \emph{Spitzer} bands. The mean spline/quadratic function ratio $\langle R \rangle$ was then defined as
\begin{equation}
\label{eq:ratio-spline-quad}
\langle R \rangle \equiv 10^{\langle |{\rm spline}_i - {\rm quad}_i| \rangle}~,
\end{equation}
where ${\rm spline}_i$ and ${\rm quad}_i$ are, respectively, the spline representation and the quadratic function evaluated on a grid of 1000 values covering the wavelength range defined by the 4 middle points, all in $(\log_{10}(\lambda), \log_{10}(F_\nu))$ space. By using the absolute difference, we make sure that a perfect agreement translates into $\langle R \rangle = 1$, and that any deviation from one curve with respect to the other in any direction would produce $\langle R \rangle > 1$.
 
\begin{figure*}[!t]
\centering
\includegraphics[width=0.99\textwidth]{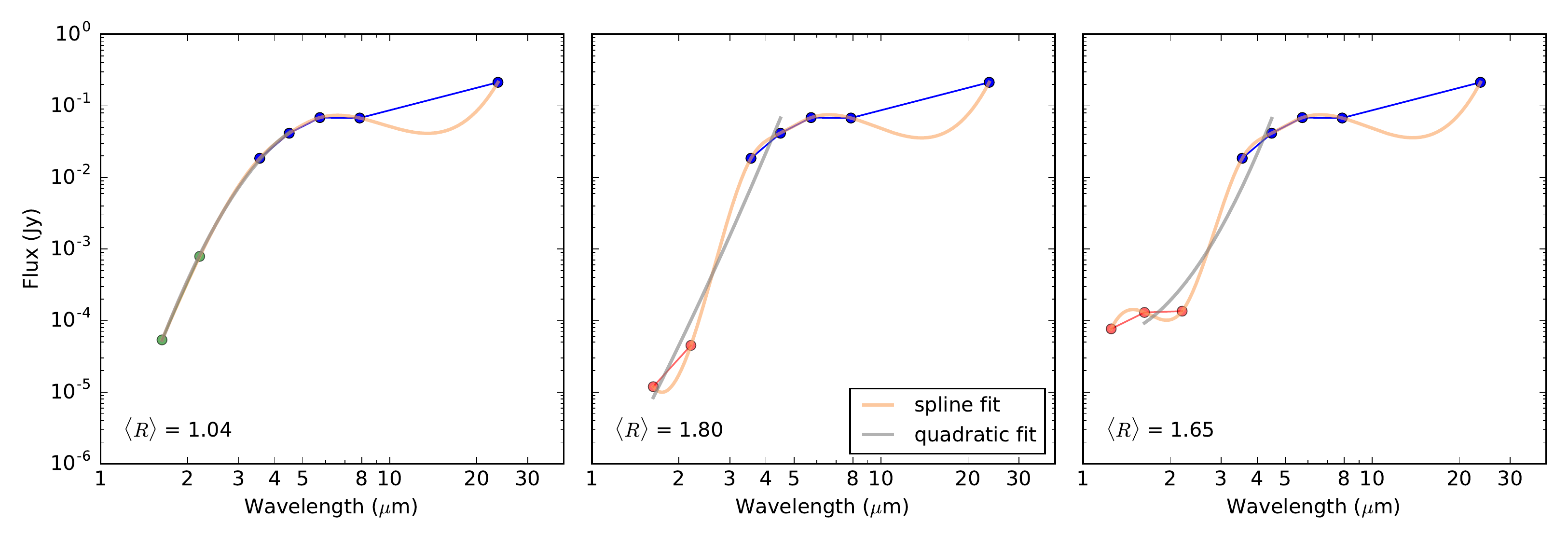}
\caption{Example of the method implemented in this work to evaluate the smoothness of the NIR-MIR transition of the SED constructed for every UKIDSS source and the corresponding \citetalias{Robitaille2008} object. Here, for each one of the three UKIDSS sources detected for the \citetalias{Robitaille2008} object SSTGLMC~G048.9610$-$00.3963 (the same shown in Figure~\ref{fig:sed-usefulness}), we plot the combined NIR-MIR SED overlaid with its cubic spline representation (yellow line) and the quadratic function fitted over the 4 middle points ($H$, $K$, 3.6~\micron, and 4.5~\micron\ filters) (gray line). The value for the mean ratio of both curves as defined in Equation~\ref{eq:ratio-spline-quad} is indicated at the lower left corner of each panel. Colors for the SED points are as in the right panel of Figure~\ref{fig:sed-usefulness}. Note that, in this particular example, the $J$ upper limits have been simply ignored by the method (see the text for details on how different cases like this are treated).}
\label{fig:sed-fit}
\end{figure*}

Figure~\ref{fig:sed-fit} illustrates this method for the same \citetalias{Robitaille2008} object previously shown in Figure~\ref{fig:sed-usefulness}. Now we plot individually the combined NIR-MIR SED for each detected UKIDSS source, together with its spline representation and the quadratic fit described above; we also indicate the value of  $\langle R \rangle$ for each source. It is clear from this example that the agreement or disagreement of both curves represents a good indicator of the match or mismatch between the NIR and MIR parts of the SED, and that this is well quantified by the mean ratio $\langle R \rangle$ defined in Equation~\ref{eq:ratio-spline-quad}.

When the flux at a given NIR wavelength is flagged as a bad quality measurement or represents an upper limit, but the other two fluxes unambiguously define the SED (mis)match, the missing flux can be simply ignored by the method, as done for the $J$ upper limits in the example shown in Figure~\ref{fig:sed-fit}. Detailed decision rules for all the possible combinations of flagged fluxes and upper limits are described in Appendix~\ref{sec:decision-rules}. In summary, we did not run the SED matching method for UKIDSS sources satisfying one or more of the following  
conditions: flagged $K$-band flux; number of flagged fluxes $>1$; or total number of flagged fluxes and upper limits $=3$. However, in some cases the associated \citetalias{Robitaille2008} object was still included in the statistics presented in Sections~\ref{sec:statistics} and \ref{sec:dominant-sources}, especially if the rejected UKIDSS source had a sufficiently low $K$-band flux. Out of the $14\,291$ detected UKIDSS sources within 2\arcsec\ of non-saturated \citetalias{Robitaille2008} objects (recall Section~\ref{sec:quality}), $12\,257$ were considered for the SED matching procedure (the remaining 2034 were rejected), of which $10\,696$ are ``unambiguous sources'' (category \verb|G| as defined in Appendix~\ref{sec:decision-rules}), i.e., they are detected in all the UKIDSS bands or present an upper limit or bad quality flux that does not compromise the unambiguity of the SED (mis)match.

\begin{figure}[!t]
\centering
\includegraphics[width=0.49\textwidth]{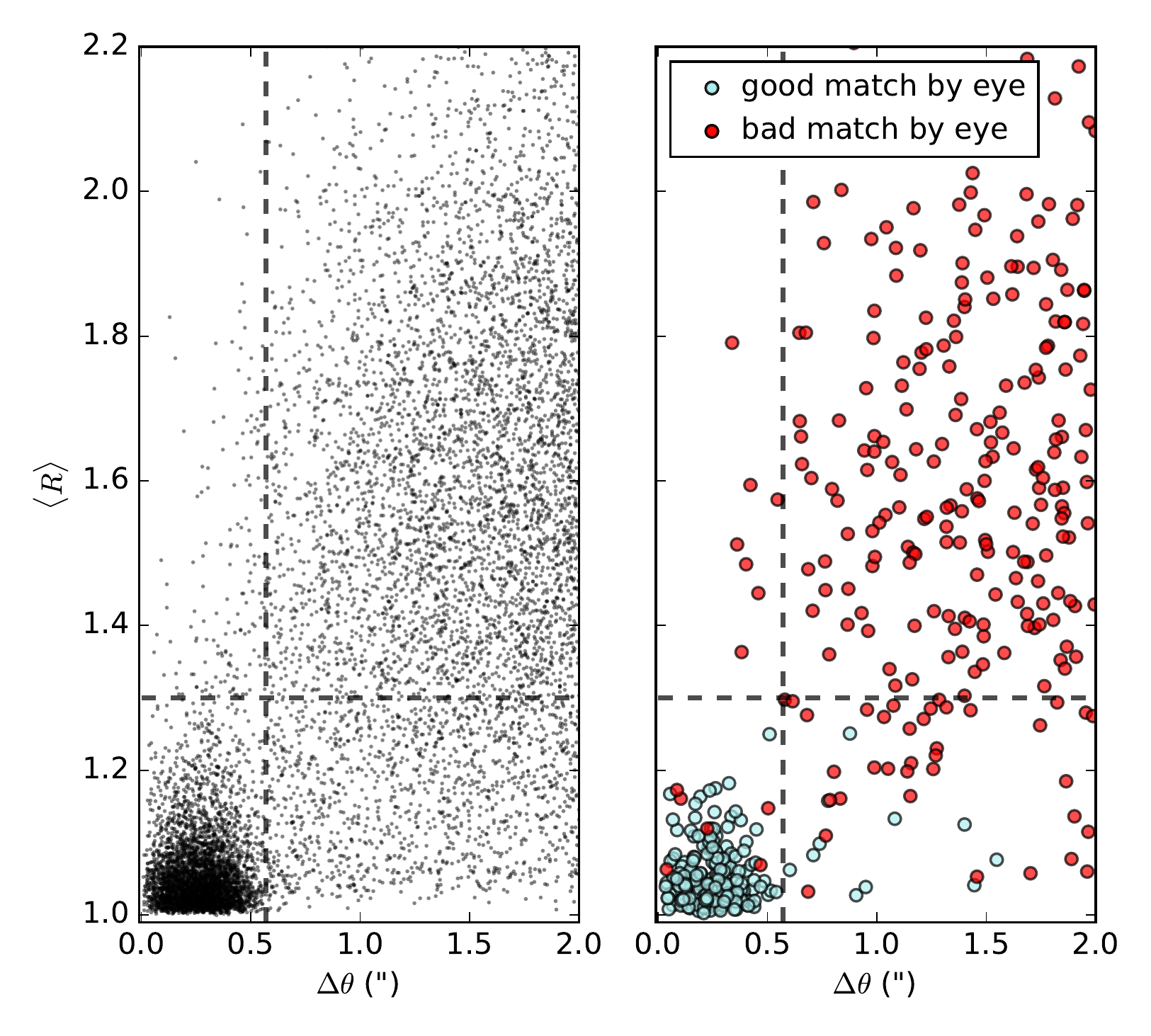}
\caption{Mean spline/quadratic function ratio $\langle R \rangle$ (as defined in Equation~\ref{eq:ratio-spline-quad}) versus the angular distance $\Delta \theta$ from the associated \citetalias{Robitaille2008} object, for all ``unambiguous'' UKIDSS sources detected (\emph{left} panel), and for the sources from the validation sample (\emph{right} panel), which are color-coded according to their visual evaluation of good (pale sky-blue) or bad (red) match with the MIR SED. The limits on $\langle R \rangle$ and $\Delta \theta$ defining the quantitative criteria of Equation~\ref{eq:sed-match-criteria} are indicated as dashed lines on both panels.} 
\label{fig:sed-match-criteria}
\end{figure}

From all the \citetalias{Robitaille2008} objects associated with this list of unambiguous UKIDSS sources, we randomly selected a control sample of 75 objects, which uniformly cover the GLIMPSE observed area for $\ell \ge -2$, and are associated with 157 UKIDSS sources. We inspected the combined SEDs of all the objects in this control sample (similar plots to the ones shown in Figure~\ref{fig:sed-fit}), and visually decided whether or not the NIR and MIR parts of the SED are well matched. We found that a very useful parameter combination in which the sources well matched by eye were clearly separated from the poorly matched ones was the 2D-space defined by $\langle R \rangle$ and the angular separation between the UKIDSS source and its corresponding \citetalias{Robitaille2008} object, hereafter denoted by $\Delta \theta$. Out of the 73 UKIDSS sources that were visually evaluated as having a good match with the MIR SED, 70 are concentrated in a confined area defined by the conditions
\begin{equation}
\label{eq:sed-match-criteria}
\begin{cases} \Delta \theta &\le \,0.57\arcsec \\
              \langle R \rangle &\le \,1.3~,
\end{cases}
\end{equation}
which represent, for the rest of this paper, our quantitative criteria to distinguish the good SED matches from the bad ones. To estimate the possible contamination of using these criteria, we selected an independent random sample of 220 uniformly distributed \citetalias{Robitaille2008} objects, associated with a total of 471 UKIDSS sources which define what we call the \emph{validation sample}. This sample was examined through the same visual inspection process we carried out for the control sample, in order to compare the visual criterion with the quantitative criteria defined by Equation~\ref{eq:sed-match-criteria}. The \emph{left} panel of Figure~\ref{fig:sed-match-criteria} shows the location of all unambiguous UKIDSS sources in the $\langle R \rangle$ vs. $\Delta \theta$ plot, while the \emph{right} panel shows the same diagram for the sources from the validation sample, which are color-coded according to their evaluation after the visual inspection. We found that 199 sources from the validation sample fell within the area delineated by Equation~\ref{eq:sed-match-criteria}, of which only 6 were visually considered as bad matches, implying a misclassification fraction of 3\%. On the other hand, 11 sources visually considered as good matches are outside that area, but notably they are all in the $\langle R \rangle \le 1.3$ region, where a total of 47 sources from the validation sample are found (with $\Delta \theta > 0.57\arcsec$). This translates into no contamination for bad matches defined by $\langle R \rangle > 1.3$, and 23\% contamination for bad matches defined by $\Delta \theta > 0.57\arcsec$ but having $\langle R \rangle \le 1.3$. These contamination fractions will be used in Sections~\ref{sec:statistics} and \ref{sec:dominant-sources} to properly estimate the uncertainties on the statistics presented there. 

We note that there is another very specific cause of contamination for sources satisfying Equation~\ref{eq:sed-match-criteria} that was not revealed by the validation sample, and was treated independently. This consists of sources for which the derived spline and the fitted quadratic function are in good agreement, but the combined SED does not follow the expected concave shape of a reddened source in the NIR, and therefore the quadratic function is convex. With the exception of very peculiar cases, these sources were considered as having uncertain match with the MIR SED. In Appendix~\ref{sec:convex}, we give more details on how we processed these sources.

\subsection{Statistics of SED match}
\label{sec:statistics}

Using the SED matching criteria defined in Equation~\ref{eq:sed-match-criteria}, we can now classify the \citetalias{Robitaille2008} objects in a way that help us to address the potential multiplicity of these objects when seen in UKIDSS, in particular the question of whether there is only one dominant contributor to the GLIMPSE fluxes (Section~\ref{sec:dominant-sources}). According to the number of detected UKIDSS sources within 2\arcsec, how many of them match the MIR SED, and whether or not one of them dominates the flux in the $K$ band (in cases when the SED (mis)match is not enough), each \citetalias{Robitaille2008} object can be in one of the following categories:

\begin{description}[labelwidth=\widthof{\textbullet\ \texttt{UM\_SM}:}, leftmargin=!]

 \item[\textbullet\ \texttt{U0}:] No UKIDSS source detected within 2\arcsec.
 \item[\textbullet\ \texttt{U1\_S1}:] Only one UKIDSS source detected, which matches the MIR SED.
 \item[\textbullet\ \texttt{U1\_S0}:] Only one UKIDSS source detected, which does not match the MIR SED.
 \item[\textbullet\ \texttt{UM\_S1}:] Multiple (more than one) UKIDSS sources detected, but only one matching the MIR SED.
 \item[\textbullet\ \texttt{UM\_SM}:] Multiple UKIDSS sources detected, more than one matching the MIR SED and with $K$-band fluxes within a factor~10.
 \item[\textbullet\ \texttt{UM\_SM\_K10}:] Multiple UKIDSS sources detected, more than one matching the MIR SED, but among them there is one source that is brighter at $K$ than the others by a factor $>10$.
 \item[\textbullet\ \texttt{UM\_S0}:] Multiple UKIDSS sources detected, none matching the MIR SED, and all having $K$-band fluxes within a factor~10.
  \item[\textbullet\ \texttt{UM\_S0\_K10}:] Multiple UKIDSS sources detected, none matching the MIR SED, but one source is the brightest at $K$ by a factor $>10$.

\end{description}

Instead of directly counting the number of \citetalias{Robitaille2008} objects that fall in each category, we run $10^5$~Monte Carlo (MC) simulations to include the presence of contaminants in the SED matching criteria of Equation~\ref{eq:sed-match-criteria}, allowing us to estimate the uncertainties on this classification. In Appendix~\ref{sec:MC-contamination}, we explain the details of this procedure.

\begin{table*}[!t]
\renewcommand{\arraystretch}{1.1}
\caption{Statistics for the classification of \citetalias{Robitaille2008} objects.}
\label{tab:classification}
\centering
\begin{tabular}{lcccccc}
\hline\hline
Category & $\langle N \rangle$ & $\sigma(N)$ & $\%_{\rm YSOs}$ & 
$\sigma(\%_{\rm YSOs})$ & $\%_{\rm AGB}$ & $\sigma(\%_{\rm AGB})$\\
%(1) & (2) & (3) & (4) & (5) & (6) & (7)\\
\hline
all objects & 5355 & \nodata\tablefootmark{a} & 71.4 & \nodata\tablefootmark{a} & 28.6 & \nodata\tablefootmark{a} \\
ambiguous/special cases\tablefootmark{b} & 939 & 17 & 75.9 & 1.4 & 24.1 & 1.4 \\
\verb|U0| & 134 & \nodata\tablefootmark{a} & 90.3 & 3.9 & 9.7 & 3.9 \\
\verb|U1_S1| & 964 & 14 & 69.6 & 1.3 & 30.4 & 1.3 \\
\verb|U1_S0| & 60 & 13 & 76.8 & 7.2 & 23.2 & 7.2 \\
\verb|UM_S1| & 2504 & 47 & 65.3 & 0.7 & 34.7 & 0.7 \\
\verb|UM_SM| & 265 & 50 & 82.0 & 3.1 & 18.0 & 3.1 \\
\verb|UM_SM_K10| & 11 & 4 & 83 & 16 & 17 & 16 \\
\verb|UM_S0| & 56 & 11 & 74.9 & 7.3 & 25.1 & 7.3 \\
\verb|UM_S0_K10| & 53 & 16 & 59.8 & 8.5 & 40.2 & 8.5 \\
low $S/N$ in the $K$-band & 370 & 5 & 91.6 & 2.3 & 8.4 & 2.3 \\
\hline
\end{tabular}
\tablefoot{
$\langle N \rangle$ is the mean number of objects that fall in each category for the whole set of $10^5$~MC simulations, while $\sigma(N)$ represents the corresponding standard deviation. The remaining columns give the fractions (in percentages) of YSOs and AGB stars for the objects within each category ($\%_{\rm YSOs}$ and $\%_{\rm AGB}$, respectively), and the estimated $1\sigma$ uncertainties on these fractions ($\sigma(\%_{\rm YSOs})$ and $\sigma(\%_{\rm AGB})$, respectively). The different categories are explained in the text. The sample of \citetalias{Robitaille2008} objects used to construct this table consists of the 5355 objects not affected by any kind of saturation in the UKIDSS images.
\tablefoottext{a}{Fixed number, not affected by MC simulations.}
\tablefoottext{b}{These particular cases are detailed in Appendix~\ref{sec:extended-sample}.}
}
\end{table*}

In Table~\ref{tab:classification}, we list the mean number of \citetalias{Robitaille2008} objects that fall in each category, and the corresponding standard deviation (Columns $\langle N \rangle$ and $\sigma(N)$, respectively), for the whole set of MC simulations. Here, we have used the sample of 5355 \citetalias{Robitaille2008} objects not affected by any kind of saturation. It is already clear from the table that the categories indicating \citetalias{Robitaille2008} objects with only one dominant UKIDSS counterpart (\verb|U1_S1| and \verb|UM_S1|) largely outnumber the remaining classes. This will be further investigated in Section~\ref{sec:dominant-sources}. The last row of Table~\ref{tab:classification} counts such objects for which the UKIDSS sources do not have a signal-to-noise $S/N$ high enough to produce an unequivocal classification. We used a threshold of $S/N = 30$ in the $K$-band, so that any undetected UKIDSS source at $K$, i.e., below $3\sigma = 3N$, would still be fainter by a factor $>10$ than the relevant UKIDSS source for each category\footnote{The ``relevant'' UKIDSS source for which we impose the threshold in $S/N$ depends on the specific category. For \texttt{U1\_S1} and \texttt{U1\_S0}, there is an unique UKIDSS source to evaluate; for \texttt{UM\_S1}, we consider the source which matches the SED; and for \texttt{UM\_S0\_K10} and \texttt{UM\_S0}, we simply evaluate the brightest $K$-band source. \texttt{UM\_SM} objects were not considered, because the eventual presence of undetected UKIDSS objects within a factor $\le 10$ at $K$ would not affect its classification. \texttt{UM\_SM\_K10} objects are treated independently in Appendix~\ref{sec:extended-sample}.}. 

We also derived the fraction of objects, within each category of Table~\ref{tab:classification}, that were classified as YSO or AGB star candidates by \citetalias{Robitaille2008} (Columns $\%_{\rm YSOs}$ and $\%_{\rm AGB}$, respectively). The uncertainty on each fraction (Column $\sigma(\%_{\rm YSOs})$ or $\sigma(\%_{\rm AGB})$) was estimated by adding in quadrature the standard deviation of the fraction computed for all the MC simulations and the typical error derived from subsampling the whole set of objects in each category (using the hypergeometric distribution; see Appendix~\ref{sec:hypergeometric} for details). Out of the 5355 \citetalias{Robitaille2008} objects not affected by any kind of saturation in UKIDSS, 1530 and 3825 are classified as AGB star and YSO candidates, respectively. Although the AGB stars/YSOs distinction made by \citetalias{Robitaille2008} is not perfect for individual objects, it provides a reasonable separations in a statistical sense (see Section~\ref{sec:glimpse-data})\footnote{In principle, the misclassification from the AGB stars/YSOs separation could be also modeled within the MC simulations, as for our SED matching criteria. Unfortunately, there is no robust determination of the contamination of using this separation that could be applied to the whole inner Galactic plane.}. The fraction of YSO and AGB star candidates in some categories of Table~\ref{tab:classification} indeed show significant (and expected) differences (at the $2\sigma$ level) with respect to the whole sample of \citetalias{Robitaille2008} objects considered (first row). Because AGB stars are generally brighter than YSOs, there is a relatively lower fraction of AGB star candidates within the \verb|U0| category and within the set of objects with low $S/N$. Similarly, given that AGB stars are more isolated than YSOs, the proportion of AGB star candidates is  higher for \verb|UM_S1| objects, and lower for \verb|UM_SM| objects (see Section~\ref{sec:dominant-sources}).
% for the last point: also YSO are more easily resolved than AGB stars?

\subsection{Fraction of GLIMPSE YSO candidates having one dominant UKIDSS counterpart}
\label{sec:dominant-sources}

In this Section, we use the classification of \citetalias{Robitaille2008} objects based on the SED match/mismatch with the detected UKIDSS sources (see Section~\ref{sec:statistics}) to estimate the fraction of YSOs that can be described by only one dominant UKIDSS counterpart. For simplicity, we will refer those objects as ``UKIDSS-single'', though there might be other non-dominant UKIDSS sources physically associated with the same object.

Since we are particularly interested in the sample of YSOs, and AGB stars are relatively more isolated and thus most of them are expected to be UKIDSS-single, here we adopt the AGB stars/YSOs separation made by \citetalias{Robitaille2008}, as in Section~\ref{sec:statistics}. For each MC simulation previously performed, we counted in each sample -- candidate YSOs and AGB stars -- the objects which fall in the SED matching categories that are consistent with being UKIDSS-single objects, versus the objects with classification consistent with not being UKIDSS-single objects. Then, we can compute the mean fraction of UKIDSS-single objects (and the standard deviation) of the whole set of MC simulations. Here, we have only used the ``good-quality'' categories, in which the UKIDSS-single nature is reliably defined: \verb|U1_S1| and \verb|UM_S1| objects for UKIDSS-single, and \verb|U1_S0|, \verb|UM_SM|, \verb|UM_SM_K10|, \verb|UM_S0| and \verb|UM_S0_K10| for non-UKIDSS-single objects, comprising a total of $\sim 1260$~AGB star candidates and $\sim 2650$~YSO candidates (exact numbers vary for each MC simulation).

The fraction of UKIDSS-single objects turned out to be $92.1 \pm 1.2 \%$ for candidate AGB stars, and $87.0 \pm 1.6 \%$ for candidate YSOs. The higher fraction of UKIDSS-single objects for AGB stars with respect to YSOs, as stated above, is expected. However, the absolute fraction of UKIDSS-single YSOs is, at first sight, surprisingly high, because one would expect a larger number of GLIMPSE YSOs with potentially multiple dominant UKIDSS counterparts, given the typically clustered nature of their environment. Possible explanations of this result will be discussed in Section~\ref{sec:synth-clustering-glimpse}. If we include in the statistics \citetalias{Robitaille2008} objects affected by peripheral saturation, as well as some special cases that could be considered as UKIDSS-single, the computed fractions are identical within the uncertainties (see Appendix~\ref{sec:extended-sample}).

We compared the angular separation $\Delta\theta$ between the UKIDSS sources and the respective \citetalias{Robitaille2008} objects which were classified as UKIDSS-single but enclosing more than one detected source in total (i.e., all \verb|UM_S1| objects). We found that almost all the dominant UKIDSS sources (99\% of the cases) were also the nearest in angular projection to the corresponding \citetalias{Robitaille2008} objects, which is consistent with the fact that these sources are the unique main contributors to the GLIMPSE fluxes. This means that, at least for the \citetalias{Robitaille2008} sample, and given that the majority of their objects have only one dominant UKIDSS counterpart, a simple ``nearest-source'' matching between UKIDSS and GLIMPSE would be a statistically good approximation. However, this kind of matching would fail for the few objects which show possibly genuine multiplicity in UKIDSS (see~\ref{sec:multiple-objects}), or for different samples of \emph{Spitzer}-selected YSO candidates that include fainter objects and would probably have a lower fraction of UKIDSS-single objects (e.g., for star-forming regions in the solar neighborhood).

\section{Discussion}
\label{sec:discussion}

\subsection{Nature of GLIMPSE objects not dominated by one UKIDSS source}
\label{sec:multiple-objects}

Even though most of the \citetalias{Robitaille2008} objects were found to be dominated by only one UKIDSS counterpart which matches the MIR SED and thus represent what we called UKIDSS-single objects (see Section~\ref{sec:dominant-sources}), it is still interesting to investigate the nature of the less frequent remaining cases. We then performed a quick visual inspection of the combined SEDs and UKIDSS images of the objects that were not labeled as UKIDSS-single, using a direct classification of the \citetalias{Robitaille2008} objects into the categories defined in Section~\ref{sec:statistics} (i.e., without MC simulations).

\begin{figure*}[!t]
\centering
\includegraphics[width=0.99\textwidth]{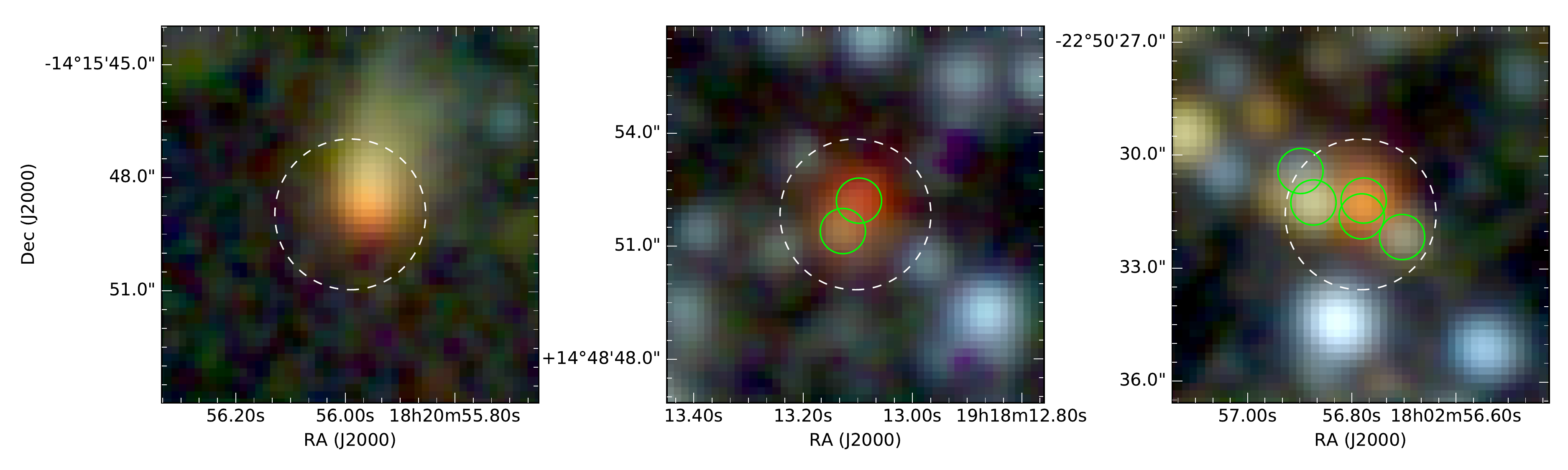}
\caption{UKIDSS $JHK$ three-color images for three examples of \citetalias{Robitaille2008} objects not dominated by only one UKIDSS source. The green circles show the positions of the detected UKIDSS sources within a radius of 2\arcsec, which is indicated by the dashed-line circle. \emph{Left:} SSTGLMC~G016.7954$+$00.1216, example of object for which the UKIDSS counterpart is too extended to be detected by the PSF-fitting photometry. \emph{Middle:} SSTGLMC~G049.1319$+$00.9327, example of object with two UKIDSS sources which match the MIR SED. \emph{Right:} SSTGLMC~G007.2266$-$00.2728, example of an apparent small cluster of UKIDSS sources.}
\label{fig:multiple-ukidss}
\end{figure*}

We found that \verb|U0| objects seem to be truly undetected in the UKIDSS images, except a few cases in which there is a UKIDSS source that is too extended to be detected by the PSF-fitting photometry (see \emph{left} panel of Figure~\ref{fig:multiple-ukidss} for an example). For most of the \verb|U1_S0| objects, the detected UKIDSS source is an unrelated star for which the SED matching criteria are correctly not satisfied (angular separation larger than the threshold, $\Delta \theta > 0.57\arcsec$, and convex quadratic fits -- see Appendix~\ref{sec:convex}). However, a few objects in this category are variable (see Section~\ref{sec:variable-sources}), or have a good SED match ($\langle R \rangle \le 1.3$, and concave quadratic fit) but their UKIDSS position is slightly shifted from the GLIMPSE position (angular separation criterion too strict in these cases).

The latter scenario can also occur for some specific UKIDSS source(s) associated with \verb|UM_S0| and \verb|UM_S0_K10| objects, especially when there seem to be genuine multiple counterparts with a noticeable angular separation which can even produce small variations in the position of the \citetalias{Robitaille2008} object in the different GLIMPSE bands (and the unique listed GLIMPSE position is separated from all the UKIDSS counterparts by more than $0.57\arcsec$). Another apparent situation for these categories is when the true UKIDSS counterpart of the \citetalias{Robitaille2008} object is below the detection limit, and the detected UKIDSS sources are unrelated objects on the line of sight. We note that, however, much of the ambiguity of all these special cases was already taken into account in our statistics of Sections~\ref{sec:statistics} and \ref{sec:dominant-sources} by the MC simulations of contamination on the SED matching criteria.

According to our visual inspection, most of the cases of apparently true multiple UKIDSS counterparts are found for \citetalias{Robitaille2008} objects classified in the \verb|UM_SM| category. Most of the \verb|UM_SM| objects have two UKIDSS sources with $K$-band fluxes within a factor~10 and that match the MIR SED (Figure~\ref{fig:multiple-ukidss}, \emph{middle}), but there are also some few cases showing three UKIDSS counterparts or even small clusters of YSOs (Figure~\ref{fig:multiple-ukidss}, \emph{right}). The factor~10 imposed here is just a nominal threshold used to separate the case of multiple UKIDSS counterparts (\verb|UM_SM|) from the situation of multiple UKIDSS sources matching the SED but one dominating the flux (\verb|UM_SM_K10|), which is very uncommon. Within the \verb|UM_SM| category, the UKIDSS counterparts typically have more similar $K$-band fluxes, with 87\% of the objects having multiple UKIDSS sources with fluxes within a factor 5.

We also found a few apparent small clusters for some \verb|UM_S0| objects, but the total number of clusters found (i.e., including those within the \verb|UM_SM| category) is not more than $\sim 15$ (we do not give an exact number here, due to the subjectivity of the visual inspection). This is consistent with the calculations presented in Section~\ref{sec:synth-clustering-glimpse}, where we found that multiple UKIDSS sources or clusters within the GLIMPSE resolution and with $K$-band fluxes within a factor~$\sim 5$ are hard to find in the \citetalias{Robitaille2008} catalog, and probably in any sample of bright single GLIMPSE YSO candidates. As an independent example, \citet{AlexanderKobulnicky2012} found only 6 UKIDSS clusters at the position of 391 bright ($<5$~mag) GLIMPSE objects selected by a red color criterion ($K_{\rm s} - [3.6] > 2$). They found additional 12 UKIDSS clusters at the positions of massive YSO candidates from the literature, which did not have a 2MASS/GLIMPSE catalog entry because of blending or saturation. Their clusters could be partially resolved at 3.6 or 4.5~\micron\ and have angular radii in the range 5\arcsec--11\arcsec, being therefore more extended than the UKIDSS clusters we found towards \citetalias{Robitaille2008} objects. However, we note that this kind of GLIMPSE objects (close to saturation, highly blended in the GLIMPSE images but missing or identified as a single entry in the catalogs) are not present in the \citetalias{Robitaille2008} catalog.

\subsection{Sensitivity of the SED match to flux changes}
\label{sec:flux-changes}

Here, we study the sensitivity of the SED match to changes in the NIR fluxes, in order to have an idea of the behavior of our method for variable sources (Section~\ref{sec:variable-sources}), as well as of the typical flux ratios between a dominant UKIDSS counterpart and fainter associated red sources for a given \citetalias{Robitaille2008} object (useful for Section~\ref{sec:synth-clustering-glimpse}). We iterated over all UKIDSS sources matching the MIR SED of non-saturated \citetalias{Robitaille2008} objects classified as \verb|U1_S1|, \verb|UM_S1|, \verb|UM_SM| and \verb|UM_SM_K10|, and gradually scaled down the UKIDSS fluxes (by an uniform fraction in all bands) to the point where the SED was not matched anymore, i.e., where $\langle R \rangle > 1.3$. We then repeated the same exercise, but now scaling up the UKIDSS fluxes.

\begin{figure}[!t]
\centering
\includegraphics[width=0.49\textwidth]{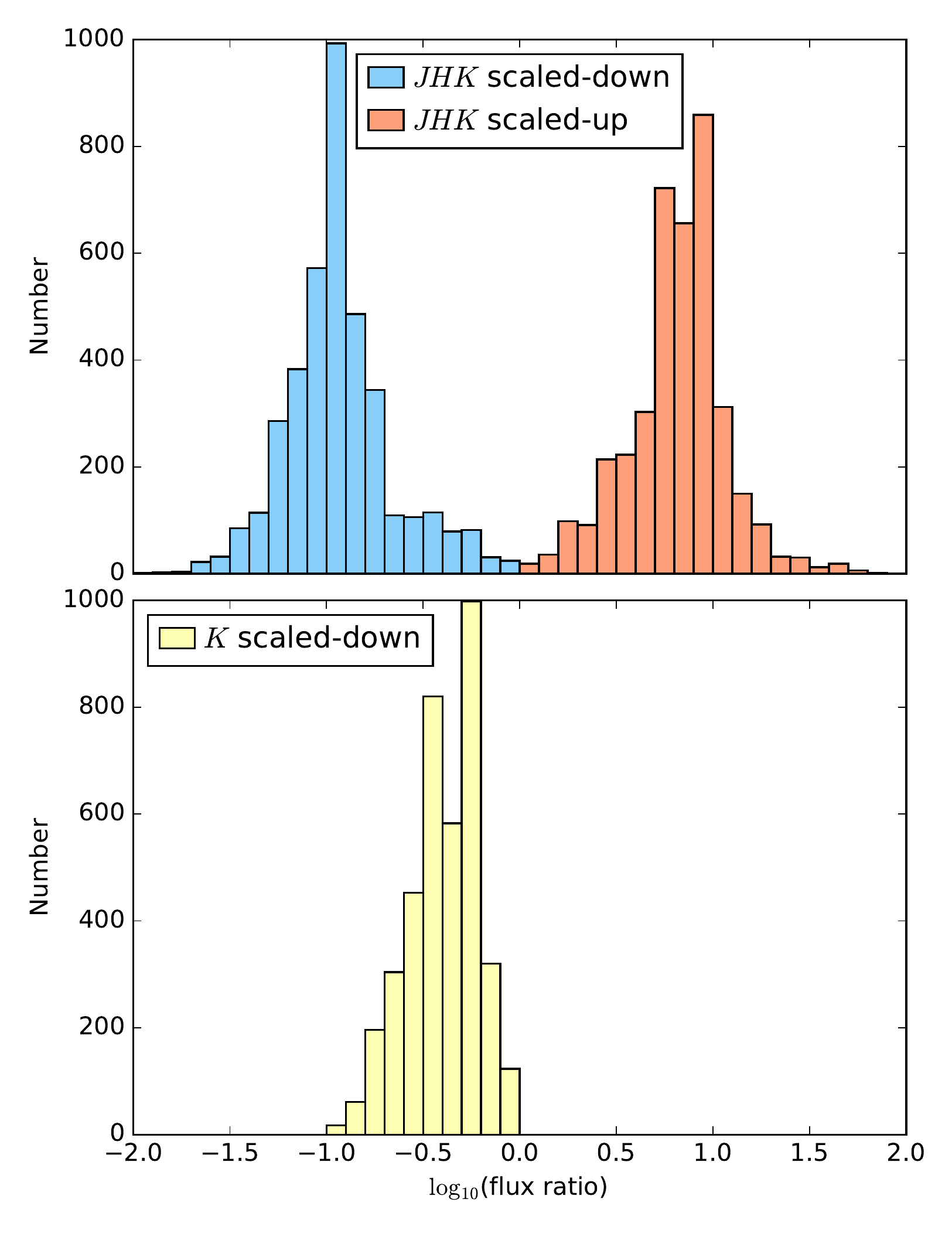}
\caption{Histograms of ratios (in logarithmic scale) between the scaled fluxes and the original ones at the point of the transition from SED match to mismatch. The top panel shows the results for fluxes that were scaled by the same amount in the three $JHK$ bands (these ratios were computed independently for scaled-up and scaled-down fluxes, and are presented here as two separated histograms). The bottom panel shows the distribution of ratios for scale-down $K$-band fluxes; in this case, the $J$ and $H$-band fluxes were kept constant.}
\label{fig:flux-ratios}
\end{figure}

The top panel of Figure~\ref{fig:flux-ratios} shows the resulting distribution of the ratio (in logarithmic scale) between the scaled fluxes and the original ones at the point of the transition from SED match to mismatch. The logarithm of the ratio have a mean and standard deviation of $-0.94 \pm 0.28$ for scaled-down fluxes, and $0.81 \pm 0.25$ for scaled-up fluxes. These values translate into an average high/low flux ratio\footnote{For convenience, note that we have inverted the ratio for scaled-down fluxes. We also remark that since our SED matching procedure was always applied in logarithmic space (see Section~\ref{sec:sed-match}), these more intuitive values are just given for reference, and the ``average'' ratio quoted here is obtained by simply inverting the mean of the logarithm computed before, so that it is not strictly the mean of the ratio.} of 8.6 and $1\sigma$ limits of $[4.6, 16.3]$ for scaled-down fluxes; and an average ratio of 6.5 and limits of $[3.7, 11.6]$ for scaled-up fluxes. This result indicates that the SED matching criteria are not very sensitive to uniform flux variations, and that fainter multiple UKIDSS counterparts could be detected (i.e., as matching the SED) if they are within a factor $\lesssim$~8--9 of the main counterpart and have a similar NIR SED shape (however, this factor could probably be closer to $\sim 5$ for real sources and not-scaled ones, see Sections~\ref{sec:multiple-objects} and \ref{sec:synth-clustering-glimpse}).

Nevertheless, this is not necessarily true for less reddened sources or background/foreground stars detected in UKIDSS. To test the performance of the SED matching method in those cases, we repeated the flux-scaling experiment described above, but this time we only scaled down the $K$-band flux, while the $J$ and $H$-band fluxes were always kept constant. In the bottom panel of Figure~\ref{fig:flux-ratios} we show the distribution of the $K$-band ratios of scaling needed for the sources to mismatch the SED. We did not scale-up the fluxes in this case, since that would only increase the $H-K$ color, which is not typical of unrelated field stars. Because the NIR SED shape of each UKIDSS source changes in this test, the resulting ratios are now quite lower, with an average high/low flux ratio of 2.4. The SED matching method is then reasonably good to reject the UKIDSS sources that are unlikely to be physically associated with a \citetalias{Robitaille2008} object. 

\subsection{Variable sources}
\label{sec:variable-sources}

Given the difference in the epochs between the GLIMPSE and UKIDSS observations, one would expect a significantly higher proportion of intrinsically variable \citetalias{Robitaille2008} objects in cases where the UKIDSS counterpart(s) do not match the MIR SED, corresponding to categories \verb|U1_S0|, \verb|UM_S0|, and \verb|UM_S0_K10| in the classification of Section~\ref{sec:statistics}. However, according to the results obtained in Section~\ref{sec:flux-changes}, the SED matching method is probably not very sensitive to variability. Indeed, by checking the subsample of objects with available GLIMPSE~II second epoch data which were found to be variable by \citetalias{Robitaille2008} by at least 0.3 mag at 4.5~\micron\ or 8.0~\micron, we did not find significant differences in the categories mentioned above. In our sample of 5355 non-saturated \citetalias{Robitaille2008} objects, 1043 have GLIMPSE~II second epoch observations, from which 151 (14.5\%) are variable. We found that only the category \verb|UM_SM| had a discrepant (at the $2\sigma$ level\footnote{The uncertainties were estimated in a completely analogous way as for YSO and AGB star fractions in each category (see Appendix~\ref{sec:hypergeometric} for details).}) proportion of variable objects with respect to the whole sample, being only 3.9\%. Though this might seem consistent, the fact that we still detected variable objects within the other categories indicating good SED match (\verb|U1_S1|, \verb|UM_S1|, and \verb|UM_SM_K10|), and with a comparable proportion to the whole sample, implies that the variability is not really discriminated by our SED matching method. 

We can also compare the UKIDSS photometry of \citetalias{Robitaille2008} objects that are dominated by one source with the corresponding 2MASS magnitudes provided in the original catalog, in order to find additional variable objects and check their classification. A total of 2083 non-saturated objects from our sample are dominated by one UKIDSS source (brightest $K$-band flux by a factor $>10$) and have 2MASS~$K_{\rm s}$ magnitudes available, which were compared with the UKIDSS~$K$ magnitudes\footnote{Here, we did not convert the 2MASS~$K_{\rm s}$ magnitudes to UKIDSS~$K$ magnitudes using the transformation given by \citet{Hodgkin2009} (as we will do later for the synthetic YSOs experiments) because for that we would need the 2MASS~$J- K_{\rm s}$ color, which would considerably reduce the comparison sample (as fewer objects have 2MASS~$J$ magnitudes available). However, the conversion is almost the identity and depends weakly on the $J- K_{\rm s}$ color, which we found to be at most $\sim 5$. The error of using directly the 2MASS~$K_{\rm s}$ magnitude is then $\lesssim 0.05$~mag.}. From this subsample, we found that 516 objects had 2MASS~$K_{\rm s}$-UKIDSS~$K$ absolute differences of more than 0.3 mag, representing 24.8\%. As expected, the fraction of $K$-variable objects is significantly higher for the candidate AGB stars ($30.8 \pm 1.2\%$) than for the YSO candidates ($21.0 \pm 0.7\%$) of this subsample. However, consistently with the results of Section~\ref{sec:flux-changes} and of GLIMPSE-variable objects, we did not obtain significant differences on the fraction of $K$-variable objects in the relevant categories of the SED matching classification, when compared with the overall fraction. The only possible trend was found when we examined the direct classification (without MC simulations) of this subsample, in which there are 5 $K$-variable objects (out of 8) within the \verb|U1_S0| category. Though the statistical significance of this trend should be confirmed by a larger sample, these variable objects might represent a small subset for which their SED shape makes the SED matching method sensitive to relatively smaller changes in the NIR fluxes; this would be  a very special situation because there are many more $K$-variable objects in the categories indicating good SED match.

\subsection{Expected clustering within the GLIMPSE resolution}
\label{sec:synth-clustering-glimpse}

In order to interpret the intuitively surprising result of Section~\ref{sec:dominant-sources} regarding the high fraction of GLIMPSE YSO candidates that are dominated by only one UKIDSS counterpart, we investigated how synthetic clustered YSOs would be observed within the GLIMPSE resolution, using the population synthesis model of Galactic YSOs developed by \citet{RobitailleWhitney2010}. The specific model used in the present work consist of a total of 2.68 million synthetic YSOs in the Galaxy, of which 9055 would be detected in GLIMPSE, included in the \citetalias{Robitaille2008} catalog, and within the YSO selection criteria adopted by \citetalias{Robitaille2008} to separate them from AGB stars\footnote{These numbers slightly differ from the comparison model used by \citet[total of 2.73 million objects, and $11\,919$ ``detected'' objects]{RobitailleWhitney2010}, since we run again a random realization of the model for a given SFR, and also because we included now the criteria of \citetalias{Robitaille2008} based on 24~\micron\ to separate the YSOs from AGB stars.}. We converted the intrinsic 2MASS~$JHK_{\rm s}$ magnitudes (i.e., before interstellar extinction is applied) to UKIDSS~$JHK$ magnitudes for all the YSOs of the model using the transformation given in Equations~(6)--(8) of \citet{Hodgkin2009}.

From the 9055 GLIMPSE-detected synthetic YSOs, we selected a subsample of 3391 objects that would be covered by UKIDSS~GPS~DR8 images (see Section~\ref{sec:ukidss-data} and Figure~\ref{fig:dr8-coverage}) and would not be saturated in UKIDSS \citep[magnitude cuts $J > 13.25$, $H > 12.75$ and $K > 12$, following][]{Lucas2008}. To simulate the effect of clustering \citep[which had not been considered by the original model by][]{RobitailleWhitney2010}, instead of rearranging the full population of synthetic YSOs into groups and clusters, we adopted the simpler approach
of placing the already selected subsample of 3391 UKIDSS-detected YSOs in clustered environments and examining how their observational properties change. For each one of these UKIDSS-detected synthetic YSOs, we randomly assigned a certain number of neighbors from the full list of 2.68 million synthetic YSOs of the model, with the only condition that they were not brighter in the $K$-band, after scaling them to the same distance and interstellar extinction.

The number of neighbors for each YSO was obtained by first drawing a value for the surface number density from a lognormal distribution, and then, using the distance, we converted this value to the number of objects within 2\arcsec\ (to compare with the properties of the UKIDSS detections in our observed sample). This assumption was motivated by the work of \citet{Bressert2010}, who found that the combined YSO surface density distribution of several \emph{Spitzer}-observed star-forming regions within 500~pc from the Sun is well described by a lognormal function. However, the \citet{Bressert2010} distribution is probably not representative of the whole range of star-forming environments in the Galaxy, in particular of the more distant and massive regions that probably harbor many of the YSO candidates from the \citetalias{Robitaille2008} sample. In fact, the Orion Nebula Cluster was excluded from the analysis by \citet{Bressert2010} due to the likely incompleteness of the \emph{Spitzer} YSO sample there, and the recent results by \citet{Kuhn2015} using YSO samples selected by combining X-ray and infrared observations towards 17 massive star-forming regions have shown that an important fraction of the YSO population in these regions resides in much denser environments than the ones studied by \citet{Bressert2010}.

\begin{figure}[!t]
\centering
\includegraphics[width=0.49\textwidth]{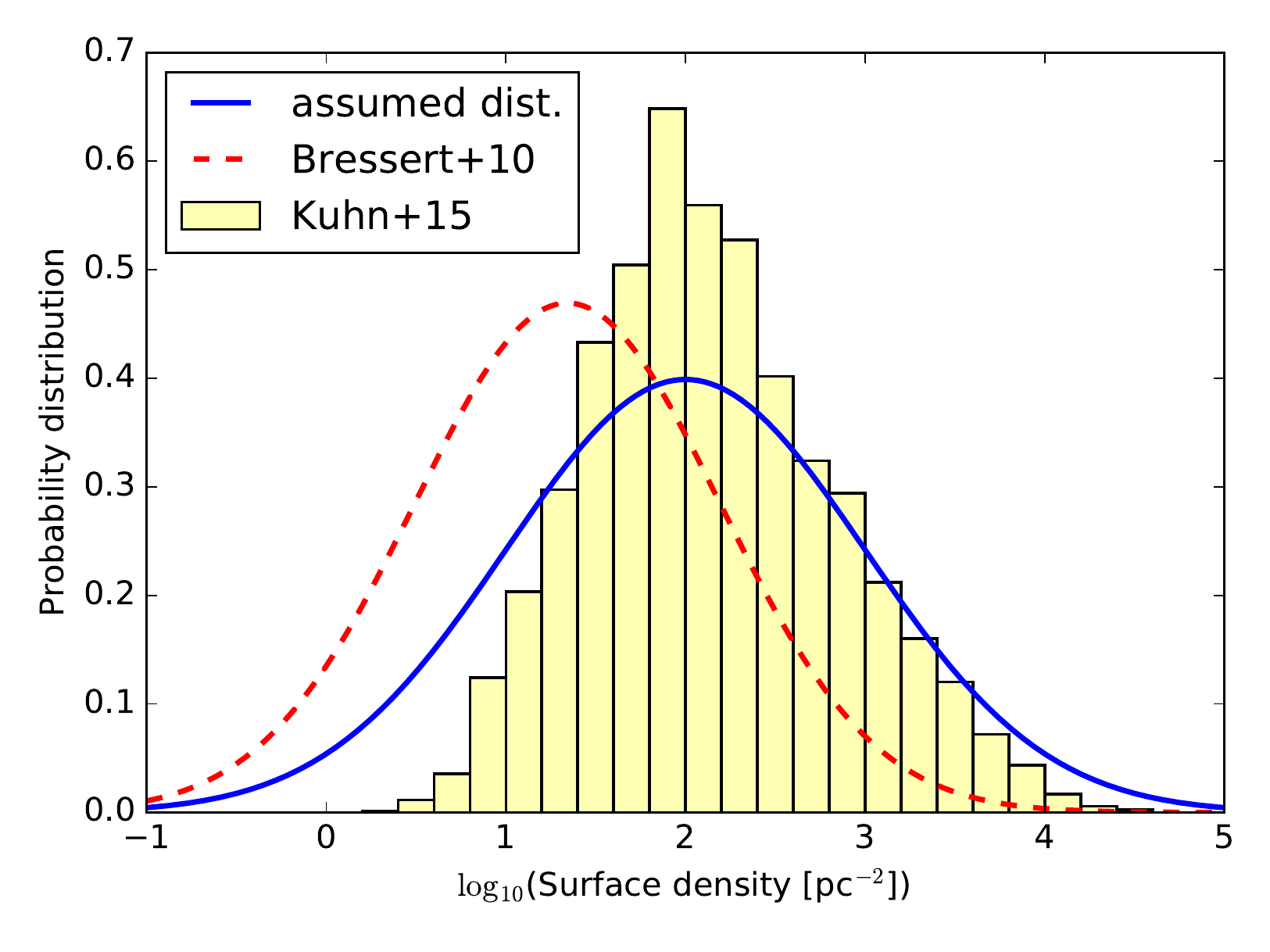}
\caption{Probability distributions of YSO surface densities, in units of pc$^{-2}$ and logarithmic scale. The solid curve represents our assumed distribution for the clustering experiments (lognormal with mean $\mu_{\log_{10}\Sigma} = 2$ and dispersion $\sigma_{\log_{10}\Sigma} = 1$), whereas the dashed curve indicates the distribution by \citet{Bressert2010} for several \emph{Spitzer}-observed star-forming regions within 500~pc from the Sun (lognormal with mean $\mu_{\log_{10}\Sigma} = 1.34$ and dispersion $\sigma_{\log_{10}\Sigma} = 0.85$). The normalized histogram is the combined surface density distribution of the 17 massive star-forming regions studied by \citet{Kuhn2015}.}
\label{fig:surface-densities}
\end{figure}

Unfortunately, there is no estimate of the average distribution of YSO surface densities in the Galaxy, which would be more appropiate for the \citetalias{Robitaille2008} sample. Here, we simply adopt a shifted and broadened lognormal distribution with respect to the one proposed by \citet{Bressert2010}: mean surface density $\mu_{\log_{10}\Sigma} = 2$ (where $\Sigma$ is in units of pc$^{-2}$), and dispersion $\sigma_{\log_{10}\Sigma} = 1$, as compared with the original $\mu_{\log_{10}\Sigma} = 1.34$ and $\sigma_{\log_{10}\Sigma} = 0.85$ of \citet{Bressert2010}. Figure~\ref{fig:surface-densities} compares these two lognormal distributions, together with the combined surface density distribution of the 17 massive star-forming regions by \citet{Kuhn2015}\footnote{We obtained this combined distribution by digitizing Figure~7 of \citet{Kuhn2015} and summing all the individual histograms.}. The assumed distribution properly covers the higher densities found by \citet{Kuhn2015}, but does not abruptly drop for the lower densities covered by \citet{Bressert2010}. Still, we believe that our adopted distribution represents an upper limit for the average clustering of YSOs in the Galaxy, which is probably somewhere in between our distribution and the one by \citet{Bressert2010}. Therefore, we also run the clustering experiments using the \citet{Bressert2010} surface density distribution as input, for comparison.

We repeated the random selection of neighbors for the sample of UKIDSS-detected synthetic YSOs in a series of 1000 MC simulations. In each simulation, we randomly assigned $K$-band detection limits (taken from the observed sample of \citetalias{Robitaille2008} objects), and counted the number of synthetic YSOs that: (1) had a signal-to-noise $S/N > 30$ in the $K$-band; (2) satisfied the $S/N$ threshold and none of their corresponding neighbors (if any) satisfy the SED matching criteria when applied to the NIR fluxes of the neighbors and the MIR fluxes of the main object (as expected, the SED matching criteria are always met when applied to the same synthetic object); or (3) satisfied the $S/N$ threshold and are brighter at $K$ than all their corresponding neighbors by at least a factor $f_K$, which was varied in the range $[4,20]$. We found that $2535 \pm 13$ UKIDSS-detected synthetic YSOs had a signal-to-noise $S/N > 30$ at $K$, of which $91.0 \pm 0.6\%$ satisfy the condition (2), and 92\% to 83\% satisfy the condition (3) for the different minimum $K$-flux ratios used. These fractions are even higher for the clustering experiments using the \citet{Bressert2010} surface density distribution: $\sim 98\%$ for condition (2), and 98\% to 95\% for condition (3).

Given the simplicity of the clustering simulations, the fractions computed for our assumed surface density distribution are in good agreement with the $87.0 \pm 1.6 \%$ of observed YSO candidates that are dominated by only one UKIDSS counterpart (referred to as UKIDSS-single objects, see Section~\ref{sec:dominant-sources}). Though we cannot discard small systematic errors derived from the assumption of our clustering experiments, the slightly lower fraction of observed UKIDSS-single objects with respect to the fraction of synthetic YSOs satisfying condition (2) might be produced by a real underestimation on the observed number of UKIDSS-single objects, which can still be present outside the categories \verb|U1_S1| and \verb|UM_S1| of the classification (see Section~\ref{sec:multiple-objects}). This hypothesis is supported by the fact that the assumed surface density distribution in our experiments probably represents an upper limit for the average YSO clustering in the Galaxy (see above).

Regarding the typical flux ratios, we found that the fraction of synthetic UKIDSS-single YSOs by SED match (condition 2) agrees with the fraction of objects satisfying condition (3) for a minimum $K$-flux ratio of $f_K = 5$, which would mean that fainter associated UKIDSS sources could be detected by the SED matching method down to a factor $1/5$ of the main counterpart. This is in agreement with the typical flux ratios found for multiple UKIDSS counterparts in the real sample (category \verb|UM_SM|, see Section~\ref{sec:multiple-objects}), but lower (although within the $1\sigma$ limits) than the average flux ratio of $f_K = 8.6$ derived from the flux-scaling experiments of Section~\ref{sec:flux-changes}, probably because in those experiments we have analyzed sources that originally matched the SED and it is therefore harder to change their SED matching criteria by just scaling down their fluxes, as compared with independent fainter sources with their own SED shape.

Overall, we have found that the high observed fraction of GLIMPSE YSO candidates that are dominated by only one UKIDSS counterpart is well reproduced by the clustering experiments presented here using a synthetic population of Galactic YSOs. Therefore, we suggest that the main explanation for this high fraction is that within the mass range covered by the \citetalias{Robitaille2008} YSO sample \citep[$\sim 3$ to $20~M_{\sun}$, see Fig.~1 by][]{RobitailleWhitney2010}, clustering with objects of comparable mass is just unlikely at the GLIMPSE resolution. However, this does not mean that clustering with lower mass YSOs does not occur, but it is just undetected by our observational method. Indeed, if we do not impose any limit on the $K$-band flux ratio, about 60\% of the UKIDSS-detected synthetic YSOs in our experiments have at least one neighbor within the GLIMPSE resolution.

\subsection{Expected clustering within the UKIDSS resolution}
\label{sec:synth-clustering-ukidss}

By doing analogous calculations to the ones presented in Section~\ref{sec:synth-clustering-glimpse}, we estimated the properties of the YSO sample when observed by a hypothetical survey with an even higher angular resolution with respect to UKIDSS. We run another set of 1000~MC experiments of random assignment of neighbors for the 3391 UKIDSS-detected YSOs (a subsample from the 9055 synthetic GLIMPSE-detected YSOs), but this time we counted the number of objects within 0.9\arcsec, the UKIDSS resolution. We found that an important fraction ($\sim 63\%$) of objects are genuinely single sources (no other sources within the UKIDSS resolution), while 97\% to 92\% have only one dominant source within the UKIDSS resolution with minimum $K$-flux ratios $f_K$ in the range $[4,20]$. This indicates that, at least for the \citetalias{Robitaille2008} sample of YSO candidates, a higher resolution than that of UKIDSS is not really needed to statistically assess their clustering properties.

We can also use the full population of 2.68 million synthetic YSOs of \citet{RobitailleWhitney2010} to generate an observed sample of YSOs that would be directly selected with UKIDSS data (without depending on GLIMPSE), and study their properties if observed by a higher resolution survey. We applied the UKIDSS magnitude cuts $13.25 < J < 19$, $12.75 < H < 18$, and $12 < K < 17$, corresponding to the saturation limits already adopted before, and rough conservative detection limits that were estimated from the UKIDSS observations of the \citetalias{Robitaille2008} objects used in this paper (see Section~\ref{sec:quality}). A 99\%-flux radius $< 0.9\arcsec$ was additionally required, following \citet{RobitailleWhitney2010}. A total of $70\,478$ synthetic YSOs satisfy these criteria within the GLIMPSE observed area (we used the full coverage in order to emulate the future inclusion of the VVV survey, see Section~\ref{sec:ukidss-data}). The experiments of random assignment of neighbors for this sample resulted in about 89\% of YSOs being genuinely single within the UKIDSS resolution, whereas 93\% to 90\% have only one dominant source within 0.9\arcsec for the same range of minimum $K$-flux ratios as before, $f_K = 4$ to $f_K = 20$. Again, we might not need a better resolution than that of UKIDSS to statistically investigate the clustering properties of a hypothetical UKIDSS-selected (and VVV-selected) sample of YSOs. We note that the objects of this sample are typically located at closer distances than the synthetic GLIMPSE-selected YSOs, as expected from the higher extinction at shorter wavelengths, so that most of the objects do not have any neighbor within 0.9\arcsec. However, since UKIDSS is more sensitive than GLIMPSE, UKIDSS-selected objects are more numerous and cover lower masses, and therefore the minimum flux ratio criterion in this case does not significantly increase the fraction of objects dominated by one source.

\subsection{Unresolved binaries}
\label{sec:binaries}

Apart from clustering, there is another characteristic of star-forming regions that can also affect results derived when assuming that GLIMPSE~YSO candidates are single objects: the ubiquitous presence of binary (and in lower proportion, multiple) systems \citep[e.g.,][]{DucheneKraus2013,Reipurth2014}, which are typically unresolved at the GLIMPSE resolution. To simulate this effect, we rearranged the 2.68 million synthetic YSOs of the model by \citet{RobitailleWhitney2010} in a certain fraction of binaries, and we then compared the number of objects (or unresolved pairs of objects) that would be selected by \citetalias{Robitaille2008}, with the 9055 selected synthetic YSOs from the original single-stars model.

There is currently a controversy on whether the binary fraction and period distribution in the Galactic field result from the dynamical processing (in dense regions or clusters) of binaries which form from an universal period distribution and a binary fraction of unity \citep[e.g.,][]{Kroupa1995,MarksKroupa2011,Marks2015}, or the binary properties in the field are indicative of the primordial binary population \citep{ParkerMeyer2014}. Therefore, we paired the synthetic YSOs in binary systems using two sets of binary properties:

\begin{itemize}[label=\textbullet]
 \item The ``universal'' setup: we assumed that all YSO are in binaries with periods that follow a universal distribution, except for massive stars. Following \citet{Oh2015}, we used the periods distributions given by their Equations (2) and (3) for primary masses $m_1 < 5 M_\sun$ and $m_1 \geq 5 M_\sun$, respectively.

 \item The ``field'' setup: we adopted primary-mass dependent binary fractions and period distributions representative of the Galactic field, summarized in Table~1 of \citet{ParkerMeyer2014}; we extended the G-dwarf range to primary masses between 0.45~$M_\sun$ and 1.5~$M_\sun$, and the A-dwarf range up to masses of 5~$M_\sun$. For primary masses $m_1 \geq 5 M_\sun$, we used the period distribution and the binary fraction of 0.69 derived by \citet{Sana2012}.
 \end{itemize}

In both setups, the secondary objects were selected in a way that the mass ratio $q = m_2/m_1$ followed a roughly flat distribution in the range $q \in [0.1,1]$, as observed in the Galactic field \citep{DucheneKraus2013, ReggianiMeyer2013}. For simplicity, the 99\%-flux radii $r_1$ and $r_2$ of the two synthetic YSOs were added geometrically to estimate the radius of the system: $r = (r_1 + r_2 + a)/2$, where $a$ is the separation of the binary, in the case when the smallest radius plus the separation is larger than the largest radius; otherwise we adopted the largest radius. Similarly, the fluxes of the two objects at every wavelength were simply added to obtain the corresponding flux of the binary system. Though this approach is probably enough to just have an idea of the effect of unresolved binaries in the observed population of GLIMPSE~YSO candidates, a proper treatment of simulated binaries would involve the use of radiative transfer models of binary YSOs instead of single YSOs for the synthetic objects, which is beyond the scope of this paper. In each setup, we counted the number of synthetic YSO (single or binary) systems that would be detected in GLIMPSE and included in the \citetalias{Robitaille2008} sample of YSO candidates (already separated from AGB stars). For binaries, we also imposed that the observed separation of the objects were outside the range $[2\arcsec,3\arcsec]$, to account for the close source criterion imposed by \citetalias{Robitaille2008} (see Section~\ref{sec:glimpse-data}).

We found that $\sim 6750$ YSO binaries from the universal setup and $\sim 7680$ YSO systems from the field setup would be selected by \citetalias{Robitaille2008} as YSO candidates\footnote{Exact numbers vary depending on each random realization. This time we did not run MC repetitions, because the pairing of a significant fraction (the totality for the universal setup) of the whole population of synthetic YSOs into binaries was much more computationally expensive than the experiments of previous sections.}, which means that, if we consider the presence of binaries, the number of GLIMPSE-selected synthetic YSOs is reduced by a factor $\sim 0.75$--0.85 with respect to the number of objects selected from a binary-free model with the same initial number of individual YSOs. The reduction factor might seem less extreme than what one could expect (e.g., $\sim 0.5$ for the case of all YSOs paired in ``twin'' binaries) just because of the pairing. Even for a flat distribution of mass ratios, primary YSOs are dominated by the more massive objects of the population, so that the number of missing massive synthetic YSOs that would have been selected by \citetalias{Robitaille2008} as single objects but are now the companions of more massive YSOs is just moderate. This can be illustrated by comparing the mass functions of the original sample and of the primary YSOs from the universal setup. The ratio of the number of primary YSOs from the universal setup with masses $m_1 > 3 M_\sun$ to the number of original single YSOs with masses $> 3 M_\sun$ (roughly all GLIMPSE-selected YSOs are above this limit) is $\sim 0.68$, which is significantly higher than 0.5, even though the total primary YSO sample is half the original.

We also used these samples of synthetic GLIMPSE-selected YSO systems from the two setups of binary properties to estimate the number of binaries that would be resolved by UKIDSS and identified by our SED matching method. The fraction of resolved or detected binaries was computed with respect to the total of GLIMPSE-selected systems (i.e., single or binary for the field setup) that were not saturated in UKIDSS and had a signal-to-noise $S/N > 30$. We assigned a $K$-band detection limit to each system, drawn from the distribution of UKIDSS sensitivities for the observed sample of \citetalias{Robitaille2008} objects (analogous to the experiments presented in Section~\ref{sec:synth-clustering-glimpse}). We found a fraction of $\sim 2\%$ to 3.5\% of binaries with observed separations larger than $0.375 \times 0.9\arcsec = 0.34\arcsec$ for the field and universal setup, respectively ($0.375\times\mathrm{FWHM}$ is the limit above which the \daophot\ routines are allowed to deblend two close sources). If we additionally required that the two objects in each binary had $K$-band fluxes within a factor of $f_K = 5$, in order to be detected by the SED matching method (see Section~\ref{sec:synth-clustering-glimpse}), and separations smaller than 0.57\arcsec\ (one of the SED matching criteria), this fraction reduces to $\sim 0.3\%$--0.6\%. Given that we have a good-quality observed sample of about 2600 GLIMPSE YSO candidates (from Section~\ref{sec:dominant-sources}), this means that we are able to detect $\sim 8$--16 binary YSOs in UKIDSS through the SED matching method. We have indeed identified many ($\sim 50$) double systems in the \verb|UM_SM| category (see Section~\ref{sec:multiple-objects}); however, with the present data we cannot distinguish the systems which represent genuine binaries from the ones that just happen to contain two YSOs within the GLIMPSE resolution due to clustering.

\subsection{Implications for SFR estimates}

By adjusting the total number of YSOs in their population synthesis model in a way that the numbers of synthetic and observed GLIMPSE-selected objects match, \citet{RobitailleWhitney2010} estimated a SFR of the Milky Way in the range 0.68--1.45~$M_\sun$~yr$^{-1}$; this range of values accounts for the uncertainty on the completeness and on the YSO selection (to separate them from AGB stars) in the observed sample. Since they assumed that the GLIMPSE-selected YSOs are \emph{individual} objects, a significant presence of unresolved clustering or binaries within the GLIMPSE resolution could in principle modify this estimate.

However, we have seen that even though clustering can be common within the GLIMPSE resolution, the YSO candidates of the \citetalias{Robitaille2008} catalog have typically intermediate masses ($\gtrsim 3~M_\sun$), and therefore dominate the observed flux with respect to their neighbors -- if the latter are randomly drawn from a canonical mass function --, which solves the question of why most of the YSO candidates analyzed in this work are dominated by only one UKIDSS counterpart. In the clustering experiments of Section~\ref{sec:synth-clustering-glimpse}, we found that $\sim 87\%$ of the synthetic YSOs are brighter at $K$ than all their corresponding neighbors (if any) by factor $> 10$, which means that if clustering were included in the \citet{RobitailleWhitney2010} model we would need anyway almost the totality of the mass of the main object to explain the observed fluxes. Consequently, the impact of clustering on the SFR estimate by \citet{RobitailleWhitney2010} is expected to be at most a few percent, and therefore, considering the uncertainties, no significant corrections are needed from this effect.

Regarding binaries, we found in Section~\ref{sec:binaries} that the number of GLIMPSE-selected YSOs is reduced by a factor $\sim 0.75$--0.85 in a population synthesis model taking into account unresolved binary YSOs, with respect to the original model. Since the model is just a random realization of the Galactic YSO population, the total number of individual YSOs scales linearly with the number of detected objects, so that the initial number of synthetic YSOs (and hence, the SFR) of the binaries model has to be increased by a factor $\sim 1.18$--1.33 to reproduce the observed number of GLIMPSE YSO candidates. Therefore, the correction from the presence of unresolved binaries to the SFR estimate of \citet{RobitailleWhitney2010} might not be negligible, and could in part make it closer to other estimates in the literature which are systematically higher \citep{ChomiukPovich2011}.

\section{Conclusions}
\label{sec:conclusions}

We have analyzed near-infrared UKIDSS observations of a sample of 8325 objects taken from the catalog of intrinsically red sources in the Galactic plane by \citet{Robitaille2008}, which were selected using the mid-infrared GLIMPSE survey. Since UKIDSS has a better angular resolution than that of GLIMPSE by a factor $> 2$, our primary aim was to investigate whether there are multiple UKIDSS sources that might all contribute to the GLIMPSE flux, or there is only one dominant UKIDSS counterpart. We did not use the published UKIDSS point source catalog which is based on aperture photometry only, and we instead performed PSF fitting photometry at the position of every GLIMPSE red source, in order to detect and properly separate the emission of multiple overlapping sources. The main results and conclusions presented in this paper are summarized as follows:

\begin{enumerate}

\item We noticed that the dominant UKIDSS sources are typically characterized by a smooth transition between their NIR SED and the MIR SED of the corresponding \citetalias{Robitaille2008} objects. We implemented a technique to automatically recognize these UKIDSS sources, which basically consisted in a comparison between two different interpolation methods at the NIR-MIR transition of the SED. This technique is very generic and could be perfectly applied for matching SEDs across gaps at other wavelengths.

\item Most of the analyzed objects from the \citetalias{Robitaille2008} sample present only one dominant UKIDSS counterpart which matches the MIR SED (what we call ``UKIDSS-single'' objects). In particular, the fraction of UKIDSS-single objects is $92.1 \pm 1.2 \%$ for candidate AGB stars, and $87.0 \pm 1.6 \%$ for candidate YSOs, using the YSO/AGB-star approximate separation of \citetalias{Robitaille2008}. While this was expected for AGB stars, it was not intuitive for YSOs given the typically clustered nature of their environment.

\item Practically the totality of the dominant UKIDSS sources were also the nearest in angular projection to the corresponding \citetalias{Robitaille2008} objects; therefore, a simple ``nearest-source'' matching between UKIDSS and GLIMPSE would be a statistically good approximation for the overall \citetalias{Robitaille2008} sample, though it would be wrong for the few specific objects showing multiple dominant sources in UKIDSS, or for different samples of \emph{Spitzer}-selected YSOs with lower fractions of UKIDSS-single objects.

\item For the few \citetalias{Robitaille2008} objects that are not dominated by one UKIDSS source, we found by visual inspection some cases of apparently true multiple UKIDSS counterparts, typically two sources, but in exceptional cases three sources or even small clusters. However, given that the SED matching method was designed to identify the dominant UKIDSS sources, these multiple sources have $K$-band fluxes within a factor $\sim 5$, while fainter counterparts are, in general, not identified by our technique.

\item We found that the SED matching method is not very sensitive to flux changes of up to a factor $\sim$~7--9, and therefore the dominant UKIDSS sources can still be identified in the SED if they are variable (considering the difference in the epochs between the GLIMPSE and UKIDSS observations).

\item By doing simple clustering experiments using the population synthesis model by \citet{RobitailleWhitney2010}, in which we randomly assign neighbors within the GLIMPSE resolution to the ``detected'' synthetic YSOs, we were able to reproduce the high fraction of GLIMPSE YSOs that are dominated by only one UKIDSS counterpart, with a $K$-flux brighter than that of their neighbors by a factor of at least $\sim 5$. We argued that within the mass range covered by \citetalias{Robitaille2008} YSO candidates ($\sim$~3--20~$M_\sun$), clustering with objects with comparable mass is unlikely at the GLIMPSE resolution.

\item We also carried out similar experiments to study the effect of unresolved binaries in the GLIMPSE YSO sample, but this time we rearranged the full initial set of synthetic YSOs of \citet{RobitailleWhitney2010} in a certain fraction of binaries and investigated how the number of ``detected'' objects change with respect to the original single-stars model. We found that this number is reduced by a factor $\sim 0.75$--0.85.

\item According to these results, we conclude that no significant corrections are needed to the SFR estimated by \citet{RobitailleWhitney2010} from the effect of YSO clustering within the GLIMPSE resolution. However, the correction derived from the presence of unresolved YSO binaries might not be negligible, and would increase the SFR estimate by a factor $\sim 1.2$--1.3.

\end{enumerate}

The SED matching method implemented in this paper turned out to be very useful to characterize the UKIDSS observations of the GLIMPSE YSO candidates, especially for the detection of the dominant counterparts. Nevertheless, as shown by the clustering experiments of synthetic YSOs, a significant fraction of the GLIMPSE YSO candidates might contain at least two physically associated UKIDSS sources within the GLIMPSE resolution, even though only one dominates the flux. The challenge for the near future is to design procedures to identify these fainter multiple sources, in order to really take advantage of the full high-resolution information of UKIDSS, and progress towards the construction of the hierarchical YSO catalog in the long term.

\begin{acknowledgements}

We thank the referee for a thorough report that helped us to improve the clarity of the paper. This work was carried out in the Max Planck Research Group \emph{Star formation throughout the Milky Way Galaxy} at the Max Planck Institute for Astronomy (MPIA). We have used observations from the \emph{Spitzer Space Telescope}, which is operated by the Jet Propulsion Laboratory, California Institute of Technology under a contract with NASA; and from the 8$^{\rm th}$ Data Release of the UKIDSS Galactic Plane Survey \citep{Lucas2008}. We thank Peter~B. Stetson for providing us the source code of the \daophot\ package. This research made use of Astropy, a community-developed core Python package for Astronomy \citep{Astropy2013}; matplotlib, a Python library for publication quality graphics \citep{Hunter2007}; and APLpy, an open-source plotting package for Python hosted at \url{http://aplpy.github.com}.

\end{acknowledgements}

\bibliography{r08-ukidss}

\begin{thebibliography}{44}
\expandafter\ifx\csname natexlab\endcsname\relax\def\natexlab#1{#1}\fi

\bibitem[{{Aguirre} {et~al.}(2011){Aguirre}, {Ginsburg}, {Dunham}, {Drosback},
  {Bally}, {Battersby}, {Bradley}, {Cyganowski}, {Dowell}, {Evans}, {Glenn},
  {Harvey}, {Rosolowsky}, {Stringfellow}, {Walawender}, \&
  {Williams}}]{Aguirre2011}
{Aguirre}, J.~E., {Ginsburg}, A.~G., {Dunham}, M.~K., {et~al.} 2011, \apjs,
  192, 4

\bibitem[{{Alexander} \& {Kobulnicky}(2012)}]{AlexanderKobulnicky2012}
{Alexander}, M.~J. \& {Kobulnicky}, H.~A. 2012, \apjl, 755, L30

\bibitem[{{Astropy Collaboration} {et~al.}(2013){Astropy Collaboration},
  {Robitaille}, {Tollerud}, {Greenfield}, {Droettboom}, {Bray}, {Aldcroft},
  {Davis}, {Ginsburg}, {Price-Whelan}, {Kerzendorf}, {Conley}, {Crighton},
  {Barbary}, {Muna}, {Ferguson}, {Grollier}, {Parikh}, {Nair}, {Unther},
  {Deil}, {Woillez}, {Conseil}, {Kramer}, {Turner}, {Singer}, {Fox}, {Weaver},
  {Zabalza}, {Edwards}, {Azalee Bostroem}, {Burke}, {Casey}, {Crawford},
  {Dencheva}, {Ely}, {Jenness}, {Labrie}, {Lim}, {Pierfederici}, {Pontzen},
  {Ptak}, {Refsdal}, {Servillat}, \& {Streicher}}]{Astropy2013}
{Astropy Collaboration}, {Robitaille}, T.~P., {Tollerud}, E.~J., {et~al.} 2013,
  \aap, 558, A33

\bibitem[{{Benjamin} {et~al.}(2003){Benjamin}, {Churchwell}, {Babler}, {Bania},
  {Clemens}, {Cohen}, {Dickey}, {Indebetouw}, {Jackson}, {Kobulnicky},
  {Lazarian}, {Marston}, {Mathis}, {Meade}, {Seager}, {Stolovy}, {Watson},
  {Whitney}, {Wolff}, \& {Wolfire}}]{Benjamin2003}
{Benjamin}, R.~A., {Churchwell}, E., {Babler}, B.~L., {et~al.} 2003, \pasp,
  115, 953

\bibitem[{{Bressert} {et~al.}(2010){Bressert}, {Bastian}, {Gutermuth},
  {Megeath}, {Allen}, {Evans}, {Rebull}, {Hatchell}, {Johnstone}, {Bourke},
  {Cieza}, {Harvey}, {Merin}, {Ray}, \& {Tothill}}]{Bressert2010}
{Bressert}, E., {Bastian}, N., {Gutermuth}, R., {et~al.} 2010, \mnras, 409, L54

\bibitem[{{Carey} {et~al.}(2009){Carey}, {Noriega-Crespo}, {Mizuno}, {Shenoy},
  {Paladini}, {Kraemer}, {Price}, {Flagey}, {Ryan}, {Ingalls}, {Kuchar},
  {Pinheiro Gon{\c c}alves}, {Indebetouw}, {Billot}, {Marleau}, {Padgett},
  {Rebull}, {Bressert}, {Ali}, {Molinari}, {Martin}, {Berriman}, {Boulanger},
  {Latter}, {Miville-Deschenes}, {Shipman}, \& {Testi}}]{Carey2009}
{Carey}, S.~J., {Noriega-Crespo}, A., {Mizuno}, D.~R., {et~al.} 2009, \pasp,
  121, 76

\bibitem[{{Casali} {et~al.}(2007){Casali}, {Adamson}, {Alves de Oliveira},
  {Almaini}, {Burch}, {Chuter}, {Elliot}, {Folger}, {Foucaud}, {Hambly},
  {Hastie}, {Henry}, {Hirst}, {Irwin}, {Ives}, {Lawrence}, {Laidlaw}, {Lee},
  {Lewis}, {Lunney}, {McLay}, {Montgomery}, {Pickup}, {Read}, {Rees}, {Robson},
  {Sekiguchi}, {Vick}, {Warren}, \& {Woodward}}]{Casali2007}
{Casali}, M., {Adamson}, A., {Alves de Oliveira}, C., {et~al.} 2007, \aap, 467,
  777

\bibitem[{{Chomiuk} \& {Povich}(2011)}]{ChomiukPovich2011}
{Chomiuk}, L. \& {Povich}, M.~S. 2011, \aj, 142, 197

\bibitem[{{Churchwell} {et~al.}(2009){Churchwell}, {Babler}, {Meade},
  {Whitney}, {Benjamin}, {Indebetouw}, {Cyganowski}, {Robitaille}, {Povich},
  {Watson}, \& {Bracker}}]{Churchwell2009}
{Churchwell}, E., {Babler}, B.~L., {Meade}, M.~R., {et~al.} 2009, \pasp, 121,
  213

\bibitem[{{Duch{\^e}ne} \& {Kraus}(2013)}]{DucheneKraus2013}
{Duch{\^e}ne}, G. \& {Kraus}, A. 2013, \araa, 51, 269

\bibitem[{{Dunham} {et~al.}(2011){Dunham}, {Robitaille}, {Evans}, {Schlingman},
  {Cyganowski}, \& {Urquhart}}]{Dunham2011}
{Dunham}, M.~K., {Robitaille}, T.~P., {Evans}, II, N.~J., {et~al.} 2011, \apj,
  731, 90

\bibitem[{{Elia} {et~al.}(2013){Elia}, {Molinari}, {Fukui}, {Schisano}, {Olmi},
  {Veneziani}, {Hayakawa}, {Pestalozzi}, {Schneider}, {Benedettini}, {di
  Giorgio}, {Ikhenaode}, {Mizuno}, {Onishi}, {Pezzuto}, {Piazzo}, {Polychroni},
  {Rygl}, {Yamamoto}, \& {Maruccia}}]{Elia2013}
{Elia}, D., {Molinari}, S., {Fukui}, Y., {et~al.} 2013, \apj, 772, 45

\bibitem[{{Fazio} {et~al.}(2004){Fazio}, {Hora}, {Allen}, {Ashby}, {Barmby},
  {Deutsch}, {Huang}, {Kleiner}, {Marengo}, {Megeath}, {Melnick}, {Pahre},
  {Patten}, {Polizotti}, {Smith}, {Taylor}, {Wang}, {Willner}, {Hoffmann},
  {Pipher}, {Forrest}, {McMurty}, {McCreight}, {McKelvey}, {McMurray}, {Koch},
  {Moseley}, {Arendt}, {Mentzell}, {Marx}, {Losch}, {Mayman}, {Eichhorn},
  {Krebs}, {Jhabvala}, {Gezari}, {Fixsen}, {Flores}, {Shakoorzadeh}, {Jungo},
  {Hakun}, {Workman}, {Karpati}, {Kichak}, {Whitley}, {Mann}, {Tollestrup},
  {Eisenhardt}, {Stern}, {Gorjian}, {Bhattacharya}, {Carey}, {Nelson},
  {Glaccum}, {Lacy}, {Lowrance}, {Laine}, {Reach}, {Stauffer}, {Surace},
  {Wilson}, {Wright}, {Hoffman}, {Domingo}, \& {Cohen}}]{Fazio2004}
{Fazio}, G.~G., {Hora}, J.~L., {Allen}, L.~E., {et~al.} 2004, \apjs, 154, 10

\bibitem[{{Hambly} {et~al.}(2008){Hambly}, {Collins}, {Cross}, {Mann}, {Read},
  {Sutorius}, {Bond}, {Bryant}, {Emerson}, {Lawrence}, {Rimoldini}, {Stewart},
  {Williams}, {Adamson}, {Hirst}, {Dye}, \& {Warren}}]{Hambly2008}
{Hambly}, N.~C., {Collins}, R.~S., {Cross}, N.~J.~G., {et~al.} 2008, \mnras,
  384, 637

\bibitem[{{Hewett} {et~al.}(2006){Hewett}, {Warren}, {Leggett}, \&
  {Hodgkin}}]{Hewett2006}
{Hewett}, P.~C., {Warren}, S.~J., {Leggett}, S.~K., \& {Hodgkin}, S.~T. 2006,
  \mnras, 367, 454

\bibitem[{{Hodgkin} {et~al.}(2009){Hodgkin}, {Irwin}, {Hewett}, \&
  {Warren}}]{Hodgkin2009}
{Hodgkin}, S.~T., {Irwin}, M.~J., {Hewett}, P.~C., \& {Warren}, S.~J. 2009,
  \mnras, 394, 675

\bibitem[{Hunter(2007)}]{Hunter2007}
Hunter, J.~D. 2007, Computing In Science \& Engineering, 9, 90

\bibitem[{{Irwin}(2010)}]{Irwin2010}
{Irwin}, M.~J. 2010, UKIRT Newsletter, 26, 14

\bibitem[{{Kroupa}(1995)}]{Kroupa1995}
{Kroupa}, P. 1995, \mnras, 277

\bibitem[{{Kuhn} {et~al.}(2015){Kuhn}, {Getman}, \& {Feigelson}}]{Kuhn2015}
{Kuhn}, M.~A., {Getman}, K.~V., \& {Feigelson}, E.~D. 2015, \apj, 802, 60

\bibitem[{{Lawrence} {et~al.}(2007){Lawrence}, {Warren}, {Almaini}, {Edge},
  {Hambly}, {Jameson}, {Lucas}, {Casali}, {Adamson}, {Dye}, {Emerson},
  {Foucaud}, {Hewett}, {Hirst}, {Hodgkin}, {Irwin}, {Lodieu}, {McMahon},
  {Simpson}, {Smail}, {Mortlock}, \& {Folger}}]{Lawrence2007}
{Lawrence}, A., {Warren}, S.~J., {Almaini}, O., {et~al.} 2007, \mnras, 379,
  1599

\bibitem[{{Longmore} {et~al.}(2011){Longmore}, {Kurtev}, {Lucas}, {Froebrich},
  {de Grijs}, {Ivanov}, {Maccarone}, {Borissova}, \& {Ker}}]{Longmore2011}
{Longmore}, A.~J., {Kurtev}, R., {Lucas}, P.~W., {et~al.} 2011, \mnras, 416,
  465

\bibitem[{{Lucas} {et~al.}(2008){Lucas}, {Hoare}, {Longmore}, {Schr{\"o}der},
  {Davis}, {Adamson}, {Bandyopadhyay}, {de Grijs}, {Smith}, {Gosling},
  {Mitchison}, {G{\'a}sp{\'a}r}, {Coe}, {Tamura}, {Parker}, {Irwin}, {Hambly},
  {Bryant}, {Collins}, {Cross}, {Evans}, {Gonzalez-Solares}, {Hodgkin},
  {Lewis}, {Read}, {Riello}, {Sutorius}, {Lawrence}, {Drew}, {Dye}, \&
  {Thompson}}]{Lucas2008}
{Lucas}, P.~W., {Hoare}, M.~G., {Longmore}, A., {et~al.} 2008, \mnras, 391, 136

\bibitem[{{Marks} {et~al.}(2015){Marks}, {Janson}, {Kroupa}, {Leigh}, \&
  {Thies}}]{Marks2015}
{Marks}, M., {Janson}, M., {Kroupa}, P., {Leigh}, N., \& {Thies}, I. 2015,
  \mnras, 452, 1014

\bibitem[{{Marks} \& {Kroupa}(2011)}]{MarksKroupa2011}
{Marks}, M. \& {Kroupa}, P. 2011, \mnras, 417, 1702

\bibitem[{{Minniti} {et~al.}(2010){Minniti}, {Lucas}, {Emerson}, {Saito},
  {Hempel}, {Pietrukowicz}, {Ahumada}, {Alonso}, {Alonso-Garcia}, {Arias},
  {Bandyopadhyay}, {Barb{\'a}}, {Barbuy}, {Bedin}, {Bica}, {Borissova},
  {Bronfman}, {Carraro}, {Catelan}, {Clari{\'a}}, {Cross}, {de Grijs},
  {D{\'e}k{\'a}ny}, {Drew}, {Fari{\~n}a}, {Feinstein}, {Fern{\'a}ndez
  Laj{\'u}s}, {Gamen}, {Geisler}, {Gieren}, {Goldman}, {Gonzalez}, {Gunthardt},
  {Gurovich}, {Hambly}, {Irwin}, {Ivanov}, {Jord{\'a}n}, {Kerins}, {Kinemuchi},
  {Kurtev}, {L{\'o}pez-Corredoira}, {Maccarone}, {Masetti}, {Merlo},
  {Messineo}, {Mirabel}, {Monaco}, {Morelli}, {Padilla}, {Palma}, {Parisi},
  {Pignata}, {Rejkuba}, {Roman-Lopes}, {Sale}, {Schreiber}, {Schr{\"o}der},
  {Smith}, {Sodr{\'e}}, {Soto}, {Tamura}, {Tappert}, {Thompson}, {Toledo},
  {Zoccali}, \& {Pietrzynski}}]{Minniti2010}
{Minniti}, D., {Lucas}, P.~W., {Emerson}, J.~P., {et~al.} 2010, \na, 15, 433

\bibitem[{{Molinari} {et~al.}(2010){Molinari}, {Swinyard}, {Bally}, {Barlow},
  {Bernard}, {Martin}, {Moore}, {Noriega-Crespo}, {Plume}, {Testi}, {Zavagno},
  {Abergel}, {Ali}, {Andr{\'e}}, {Baluteau}, {Benedettini}, {Bern{\'e}},
  {Billot}, {Blommaert}, {Bontemps}, {Boulanger}, {Brand}, {Brunt}, {Burton},
  {Campeggio}, {Carey}, {Caselli}, {Cesaroni}, {Cernicharo}, {Chakrabarti},
  {Chrysostomou}, {Codella}, {Cohen}, {Compiegne}, {Davis}, {de Bernardis}, {de
  Gasperis}, {Di Francesco}, {di Giorgio}, {Elia}, {Faustini}, {Fischera},
  {Fukui}, {Fuller}, {Ganga}, {Garcia-Lario}, {Giard}, {Giardino}, {Glenn},
  {Goldsmith}, {Griffin}, {Hoare}, {Huang}, {Jiang}, {Joblin}, {Joncas},
  {Juvela}, {Kirk}, {Lagache}, {Li}, {Lim}, {Lord}, {Lucas}, {Maiolo},
  {Marengo}, {Marshall}, {Masi}, {Massi}, {Matsuura}, {Meny}, {Minier},
  {Miville-Desch{\^e}nes}, {Montier}, {Motte}, {M{\"u}ller}, {Natoli}, {Neves},
  {Olmi}, {Paladini}, {Paradis}, {Pestalozzi}, {Pezzuto}, {Piacentini},
  {Pomar{\`e}s}, {Popescu}, {Reach}, {Richer}, {Ristorcelli}, {Roy}, {Royer},
  {Russeil}, {Saraceno}, {Sauvage}, {Schilke}, {Schneider-Bontemps},
  {Schuller}, {Schultz}, {Shepherd}, {Sibthorpe}, {Smith}, {Smith},
  {Spinoglio}, {Stamatellos}, {Strafella}, {Stringfellow}, {Sturm}, {Taylor},
  {Thompson}, {Tuffs}, {Umana}, {Valenziano}, {Vavrek}, {Viti}, {Waelkens},
  {Ward-Thompson}, {White}, {Wyrowski}, {Yorke}, \& {Zhang}}]{Molinari2010}
{Molinari}, S., {Swinyard}, B., {Bally}, J., {et~al.} 2010, \pasp, 122, 314

\bibitem[{{Oh} {et~al.}(2015){Oh}, {Kroupa}, \& {Pflamm-Altenburg}}]{Oh2015}
{Oh}, S., {Kroupa}, P., \& {Pflamm-Altenburg}, J. 2015, \apj, 805, 92

\bibitem[{{Parker} \& {Meyer}(2014)}]{ParkerMeyer2014}
{Parker}, R.~J. \& {Meyer}, M.~R. 2014, \mnras, 442, 3722

\bibitem[{{Reggiani} \& {Meyer}(2013)}]{ReggianiMeyer2013}
{Reggiani}, M. \& {Meyer}, M.~R. 2013, \aap, 553, A124

\bibitem[{{Reipurth} {et~al.}(2014){Reipurth}, {Clarke}, {Boss}, {Goodwin},
  {Rodr{\'{\i}}guez}, {Stassun}, {Tokovinin}, \& {Zinnecker}}]{Reipurth2014}
{Reipurth}, B., {Clarke}, C.~J., {Boss}, A.~P., {et~al.} 2014, Protostars and
  Planets VI, 267

\bibitem[{{Robitaille} {et~al.}(2008){Robitaille}, {Meade}, {Babler},
  {Whitney}, {Johnston}, {Indebetouw}, {Cohen}, {Povich}, {Sewilo}, {Benjamin},
  \& {Churchwell}}]{Robitaille2008}
{Robitaille}, T.~P., {Meade}, M.~R., {Babler}, B.~L., {et~al.} 2008, \aj, 136,
  2413

\bibitem[{{Robitaille} \& {Whitney}(2010)}]{RobitailleWhitney2010}
{Robitaille}, T.~P. \& {Whitney}, B.~A. 2010, \apjl, 710, L11

\bibitem[{{Sana} {et~al.}(2012){Sana}, {de Mink}, {de Koter}, {Langer},
  {Evans}, {Gieles}, {Gosset}, {Izzard}, {Le Bouquin}, \&
  {Schneider}}]{Sana2012}
{Sana}, H., {de Mink}, S.~E., {de Koter}, A., {et~al.} 2012, Science, 337, 444

\bibitem[{{Schuller} {et~al.}(2009){Schuller}, {Menten}, {Contreras},
  {Wyrowski}, {Schilke}, {Bronfman}, {Henning}, {Walmsley}, {Beuther},
  {Bontemps}, {Cesaroni}, {Deharveng}, {Garay}, {Herpin}, {Lefloch}, {Linz},
  {Mardones}, {Minier}, {Molinari}, {Motte}, {Nyman}, {Reveret}, {Risacher},
  {Russeil}, {Schneider}, {Testi}, {Troost}, {Vasyunina}, {Wienen}, {Zavagno},
  {Kovacs}, {Kreysa}, {Siringo}, \& {Wei{\ss}}}]{Schuller2009}
{Schuller}, F., {Menten}, K.~M., {Contreras}, Y., {et~al.} 2009, \aap, 504, 415

\bibitem[{{Skrutskie} {et~al.}(2006){Skrutskie}, {Cutri}, {Stiening},
  {Weinberg}, {Schneider}, {Carpenter}, {Beichman}, {Capps}, {Chester},
  {Elias}, {Huchra}, {Liebert}, {Lonsdale}, {Monet}, {Price}, {Seitzer},
  {Jarrett}, {Kirkpatrick}, {Gizis}, {Howard}, {Evans}, {Fowler}, {Fullmer},
  {Hurt}, {Light}, {Kopan}, {Marsh}, {McCallon}, {Tam}, {Van Dyk}, \&
  {Wheelock}}]{Skrutskie2006}
{Skrutskie}, M.~F., {Cutri}, R.~M., {Stiening}, R., {et~al.} 2006, \aj, 131,
  1163

\bibitem[{{Stead} \& {Hoare}(2011)}]{SteadHoare2011}
{Stead}, J.~J. \& {Hoare}, M.~G. 2011, \mnras, 418, 2219

\bibitem[{{Stetson}(1987)}]{Stetson1987}
{Stetson}, P.~B. 1987, \pasp, 99, 191

\bibitem[{{Stetson}(1994)}]{Stetson1994}
{Stetson}, P.~B. 1994, \pasp, 106, 250

\bibitem[{{Stetson} {et~al.}(2003){Stetson}, {Bruntt}, \&
  {Grundahl}}]{Stetson2003}
{Stetson}, P.~B., {Bruntt}, H., \& {Grundahl}, F. 2003, \pasp, 115, 413

\bibitem[{{Tackenberg} {et~al.}(2012){Tackenberg}, {Beuther}, {Henning},
  {Schuller}, {Wienen}, {Motte}, {Wyrowski}, {Bontemps}, {Bronfman}, {Menten},
  {Testi}, \& {Lefloch}}]{Tackenberg2012}
{Tackenberg}, J., {Beuther}, H., {Henning}, T., {et~al.} 2012, \aap, 540, A113

\bibitem[{{Urquhart} {et~al.}(2014){Urquhart}, {Moore}, {Csengeri}, {Wyrowski},
  {Schuller}, {Hoare}, {Lumsden}, {Mottram}, {Thompson}, {Menten}, {Walmsley},
  {Bronfman}, {Pfalzner}, {K{\"o}nig}, \& {Wienen}}]{Urquhart2014}
{Urquhart}, J.~S., {Moore}, T.~J.~T., {Csengeri}, T., {et~al.} 2014, \mnras,
  443, 1555

\bibitem[{{Veneziani} {et~al.}(2013){Veneziani}, {Elia}, {Noriega-Crespo},
  {Paladini}, {Carey}, {Faimali}, {Molinari}, {Pestalozzi}, {Piacentini},
  {Schisano}, \& {Tibbs}}]{Veneziani2013}
{Veneziani}, M., {Elia}, D., {Noriega-Crespo}, A., {et~al.} 2013, \aap, 549,
  A130

\bibitem[{{Werner} {et~al.}(2004){Werner}, {Roellig}, {Low}, {Rieke}, {Rieke},
  {Hoffmann}, {Young}, {Houck}, {Brandl}, {Fazio}, {Hora}, {Gehrz}, {Helou},
  {Soifer}, {Stauffer}, {Keene}, {Eisenhardt}, {Gallagher}, {Gautier}, {Irace},
  {Lawrence}, {Simmons}, {Van Cleve}, {Jura}, {Wright}, \&
  {Cruikshank}}]{Werner2004}
{Werner}, M.~W., {Roellig}, T.~L., {Low}, F.~J., {et~al.} 2004, \apjs, 154, 1

\end{thebibliography}

\newpage

\Online

\begin{appendix}
\section{Decision rules for SED matching}
\label{sec:decision-rules}

\begin{table*}[!t]
\renewcommand{\arraystretch}{1.1}
\caption{Categories of $JHK$ combinations of good-quality (\checkmark), bad quality ($\times$) and upper limit ($\downarrow$) fluxes.}
\label{tab:decision-rules}
\centering
\begin{tabular}{c|ccc|l}
\hline\hline
Category & $J$ & $H$ & $K$ & Common property\\
\hline
\verb|NS|  & $\downarrow$ & $\downarrow$ & $\downarrow$ & not significant source\\
\hline
\verb|BK|  & $\times$     & $\times$     & $\times$     & bad quality $K$ flux\\
           & \checkmark   & $\times$     & $\times$     & \\
           & $\downarrow$ & $\times$     & $\times$     & \\
           & $\times$     & \checkmark   & $\times$     & \\
           & $\times$     & $\downarrow$ & $\times$     & \\
           & \checkmark   & \checkmark   & $\times$     & \\
           & \checkmark   & $\downarrow$ & $\times$     & \\
           & $\downarrow$ & \checkmark   & $\times$     & \\
           & $\downarrow$ & $\downarrow$ & $\times$     & \\
\hline
\verb|KU-B|& $\times$     & $\times$     & $\downarrow$ & $K$ upper limit + 2 bad or upper limits\\
           & $\times$     & $\downarrow$ & $\downarrow$ & \\
           & $\downarrow$ & $\times$     & $\downarrow$ & \\
\hline
\verb|KU-G|& $\times$     & \checkmark   & $\downarrow$ & $K$ upper limit + at least one good flux\\
           & \checkmark   & $\times$     & $\downarrow$ & \\
           & \checkmark   & $\downarrow$ & $\downarrow$ & \\
           & $\downarrow$ & \checkmark   & $\downarrow$ & \\
           & \checkmark   & \checkmark   & $\downarrow$ & \\
\hline
\verb|KO-2|& $\times$     & $\times$     & \checkmark   & $K$ only + 2 bad quality fluxes\\
\hline
\verb|KO-U|& $\times$     & $\downarrow$ & \checkmark   & $K$ only + at least one upper limit\\
           & $\downarrow$ & $\times$     & \checkmark   & \\
           & $\downarrow$ & $\downarrow$ & \checkmark   & \\
\hline
\verb|G|   & $\times$     & \checkmark   & \checkmark   & unambiguous sources (good sources)\\
           & $\downarrow$ & \checkmark   & \checkmark   & \\
           & \checkmark   & $\times$     & \checkmark   & \\
           & \checkmark   & $\downarrow$ & \checkmark   & \\
           & \checkmark   & \checkmark   & \checkmark   & \\
\hline
\end{tabular}
\tablefoot{
Bad-quality and upper limit fluxes are defined in Section~\ref{sec:quality}, whereas good-quality fluxes are simply all the remaining measurements. A specific category represents a set of combinations for which we applied the same decision rule for SED matching.}
\end{table*}

Here, we describe in detail how we treated the three UKIDSS $JHK$ bands for SED matching in the presence of bad quality fluxes or/and upper limits, as defined in Section~\ref{sec:quality}. If we consider that each UKIDSS band can be either a normal measurement (good-quality flux), a bad quality flux, or an upper limit, there are then 27 possible cases of $JHK$ combinations for each source, which were grouped in 7 different categories. A specific category is a set of combinations sharing some relevant property and for which we applied the same decision rule for SED matching. The $JHK$ combinations and the defined categories are listed in Table~\ref{tab:decision-rules}. With the exception of very specific situations that are described below, most of the combinations outside the \verb|G| category represent cases in which the match or mismatch with the MIR SED is uncertain. These UKIDSS sources are referred in this paper as \emph{ambiguous} sources, and are divided into faint ambiguous (\verb|FA|) and bright ambiguous (\verb|BA|) sources depending on their $K$-band flux, $F_K^{\rm amb}$, and the flux $F_K^{\rm match}$ of the brightest UKIDSS source associated with the same \citetalias{Robitaille2008} object and matching the MIR SED. In general (unless explicitly mentioned below when describing each category), we used the criteria:
\begin{equation}
\label{eq:FA-BA separation}
\text{ambiguous source =}\begin{cases} \verb|FA| & \text{if } F_K^{\rm amb} < F_K^{\rm match}/10 \\
                                       \verb|BA| & \text{otherwise.}
\end{cases}
\end{equation}

Note that, by definition, \citetalias{Robitaille2008} objects classified as cases \verb|UM_S0_K10| and \verb|UM_S0| in Section~\ref{sec:statistics} do not have any source matching the MIR SED, so if ambiguous sources are present, they are automatically labeled as \verb|BA| (by setting $F_K^{\rm match} = 0$). Similarly, an ambiguous UKIDSS source that is the only detected source for a given \citetalias{Robitaille2008} object is always classified as \verb|BA|, except the special situations stated below. In this way, all \citetalias{Robitaille2008} objects having \verb|FA| sources are still usable for the SED matching statistics.

A \citetalias{Robitaille2008} object having a \verb|BA| source is removed from the sample for further analysis, unless the object corresponds to the category \verb|UM_SM| of SED matching and the \verb|BA| source is not brighter than $10 F_K^{\rm match}$, in which condition the \verb|UM_SM| classification is not affected. If the category \verb|UM_SM_K10| is not considered as a case of one dominant UKIDSS counterpart (as in the statistics of Section~\ref{sec:dominant-sources}), \verb|UM_SM_K10| objects with \verb|BA| sources (as before, not brighter than $10 F_K^{\rm match}$) are also allowed, due to the following argument: if the \verb|BA| source had, in reality, a good SED match, the category would only change to \verb|UM_SM|, and the total number of \citetalias{Robitaille2008} objects not having only one dominant UKIDSS counterpart would be the same.

Below, we explain every category of $JHK$ combinations and the corresponding decision rule we applied. As before, we represent the $K$-band flux of the brightest UKIDSS source matching the MIR SED as $F_K^{\rm match}$, while the flux of the considered (potentially ambiguous) UKIDSS source is just denoted by $F_K$.

\begin{itemize}[label=\textbullet]

 \item \verb|NS|: These are very few sources that were initially detected by the source finding algorithm of \daophot, but they turned out to be below the more rigorous 3$\sigma$-detection limits defined in Section~\ref{sec:quality}, in all the UKIDSS bands. They can therefore be considered as spurious sources.

 \item \verb|BK|: Given that the $K$ band is the closest in wavelength to the GLIMPSE filters, and is therefore crucial to define the NIR-MIR transition of the SED, in this work we adopted the conservative approach of rejecting all UKIDSS sources with bad quality $K$ flux for the SED analysis. We applied for the associated \citetalias{Robitaille2008} object the rule:
 \begin{equation*}
 \begin{cases} \text{if } F_K < F_K^{\rm match}/10 :& \text{\citetalias{Robitaille2008} object with } \verb|FA| \text{ source}\\
               \text{else:} & \text{remove the object from the sample.}
 \end{cases}
 \end{equation*}
 Note that, in this particular case, even \citetalias{Robitaille2008} objects with SED matching types \verb|UM_SM| (and \verb|UM_SM_K10|) were not considered, because bright bad quality $K$ fluxes could also affect the photometry of nearby sources.

 \item \verb|KU-B|: In this case, there is no good-quality flux in any band so that the SED (mis)match is completely uncertain; however, since the $K$-band is just upper limit and not a bad quality measurement, the \citetalias{Robitaille2008} object follows the standard decision rule of Equation~\ref{eq:FA-BA separation}:
 \begin{equation*}
 \begin{cases} \text{if } F_K < F_K^{\rm match}/10 :& \text{object with } \verb|FA| \text{ source}\\
               \text{else:} & \text{object with } \verb|BA| \text{ source.}
 \end{cases}
 \end{equation*}
 
 \item \verb|KU-G|: These sources have also a $K$-band upper limit, but the presence of at least one good-quality flux in the other bands allows us to run the SED matching procedure using the upper limits (except the case of detection at $H$ only, for which we ignored the $J$ upper limit) and ignoring the bad quality fluxes. With this limited information, if the source does match the SED, this would still be uncertain because the actual $K$-band flux could be lower; however, if the source does not match the SED, this is very likely the case since the lower actual fluxes at longer wavelengths ($K$-band, and $H$-band when it is also an upper limit) would produce an even clearer SED mismatch. Then, we applied the rule:
 \begin{equation*}
 \begin{cases} \text{if source matches the SED:} & \text{equivalent to } \verb|KU-B|\\
               \text{else:} & \text{unambiguous source.}
 \end{cases}
 \end{equation*}

 \item \verb|KO-2|: Sources with good-quality flux only in the $K$-band could still be important counterparts of the respective \citetalias{Robitaille2008} objects, and therefore were not always considered ambiguous. In this particular case, if the source had a flux $F_K$ that was the brightest by a factor $>10$ among the detected sources, it was evaluated by a secondary SED matching criterion: if $\Delta \theta \le 0.57\arcsec$ (as in Equation~\ref{eq:sed-match-criteria}), and the linear extrapolation at the $K$-band from the two shortest-wavelength \emph{Spitzer} fluxes of the corresponding \citetalias{Robitaille2008} object, in $(\log_{10}(\lambda), \log_{10}(F_\nu))$ space, is within a factor 2 of $F_K$, the source is assumed to match the MIR SED; we refer this case as a ``linear'' match with the SED. If the source is not the brightest, it is treated as a standard ambiguous source. In summary:
 %, we applied the following rule for the respective \citetalias{Robitaille2008} object:
 \begin{equation*}
 \begin{cases} \text{if } F_K < F_K^{\rm match}/10 : & \verb|FA| \text{ source} \\
               \text{else if } F_K \text{ brightest by a factor} >10: & \text{evaluate} \\
                & \text{linear match.}\\ 
               \text{else:} & \verb|BA| \text{ source.} 
 \end{cases}
 \end{equation*}

 \item \verb|KO-U|: These sources have also good-quality flux in the $K$-band only, but in this case there is at least one upper limit in the other bands, so that we can run the primary SED matching method using the $K$-band flux and the upper limit(s), and ignoring any bad quality flux, if present. Given the expected shape of the SED, if the source matches the MIR SED, we think that it would likely match the SED also with the actual (lower) flux(es) at $H$ and/or $J$; if there is a SED mismatch, however, it remains uncertain, unless the source is the brightest at $K$ by a factor $>10$ and linearly matches the MIR SED, in which case the mismatch with the primary method is probably only due to the use of upper limit(s) in the other bands instead of the actual fluxes. In other words, sources that do not match the SED with the primary method were treated as \verb|KO-2| sources:
 \begin{equation*}
 \begin{cases} \text{if source matches the SED:} & \text{unambiguous source} \\
               \text{else:} & \text{equivalent to } \verb|KO-2|.
 \end{cases}
 \end{equation*}

 \item \verb|G|: This category groups all sources that have always a good-quality $K$-band flux, and at least one more good-quality flux in the $H$- and/or $J$-band. Consequently, the SED (mis)match is unambiguously defined, and the primary SED matching algorithm can be applied by simply ignoring the bad quality flux or upper limit, if present.

\end{itemize}

\section{Convex quadratic fits}
\label{sec:convex}

\begin{figure}[!ht]
\centering
\includegraphics[width=0.49\textwidth]{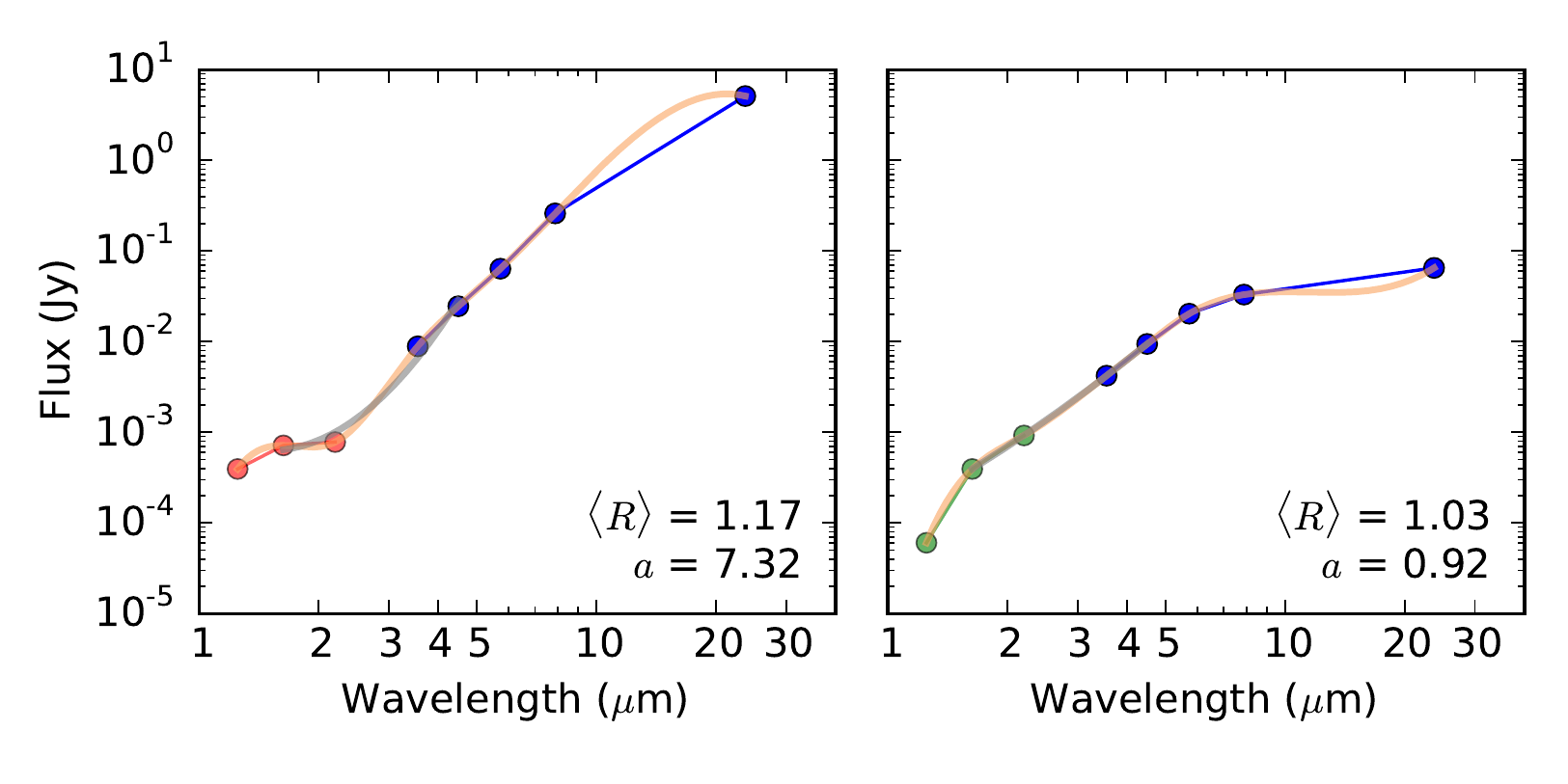}
\caption{Examples of UKIDSS sources producing a convex quadratic fit, but still satisfying the SED matching criteria of Equation~\ref{eq:sed-match-criteria}. \emph{Left:} UKIDSS source with a flat NIR SED (and therefore most likely not being a counterpart of the corresponding \citetalias{Robitaille2008} object) and high convexity. \emph{Right:} Reddened UKIDSS source that is probably the dominant counterpart of the \citetalias{Robitaille2008} object; the slight convexity of its quadratic fit is probably produced by uncertainties or variability on the measured fluxes. The values of $\langle R \rangle$ and the quadratic coefficient $a$ for each source are indicated at the lower right corner of each panel.} 
\label{fig:example-convex-sed}
\end{figure}

In this work, we have used the agreement -- quantified as the parameter $\langle R \rangle$ -- between the spline representation of the combined NIR-MIR SED and the quadratic function fitted over the 4 middle points of the SED defining the NIR-MIR transition, as one of the main indicators to distinguish the dominant UKIDSS counterparts of the \citetalias{Robitaille2008} objects. Nevertheless, in a few cases, some UKIDSS sources with a flat or blue shape in the NIR SED, and thus most likely not being a counterpart of a \citetalias{Robitaille2008} object, do succeed to produce a quadratic fit consistent with the derived spline curve and could fall within the good SED match criteria of Equation~\ref{eq:sed-match-criteria}. In such a case (see \emph{left} panel of Figure~\ref{fig:example-convex-sed} for an example), the overall SED does not follow the expected concave shape of an YSO or AGB star, and the fitted quadratic curve is convex, i.e., the coefficient $a$ of the function $f(x) = ax^2 + bx + c$ is greater than zero.

\begin{figure}[!t]
\centering
\includegraphics[width=0.49\textwidth]{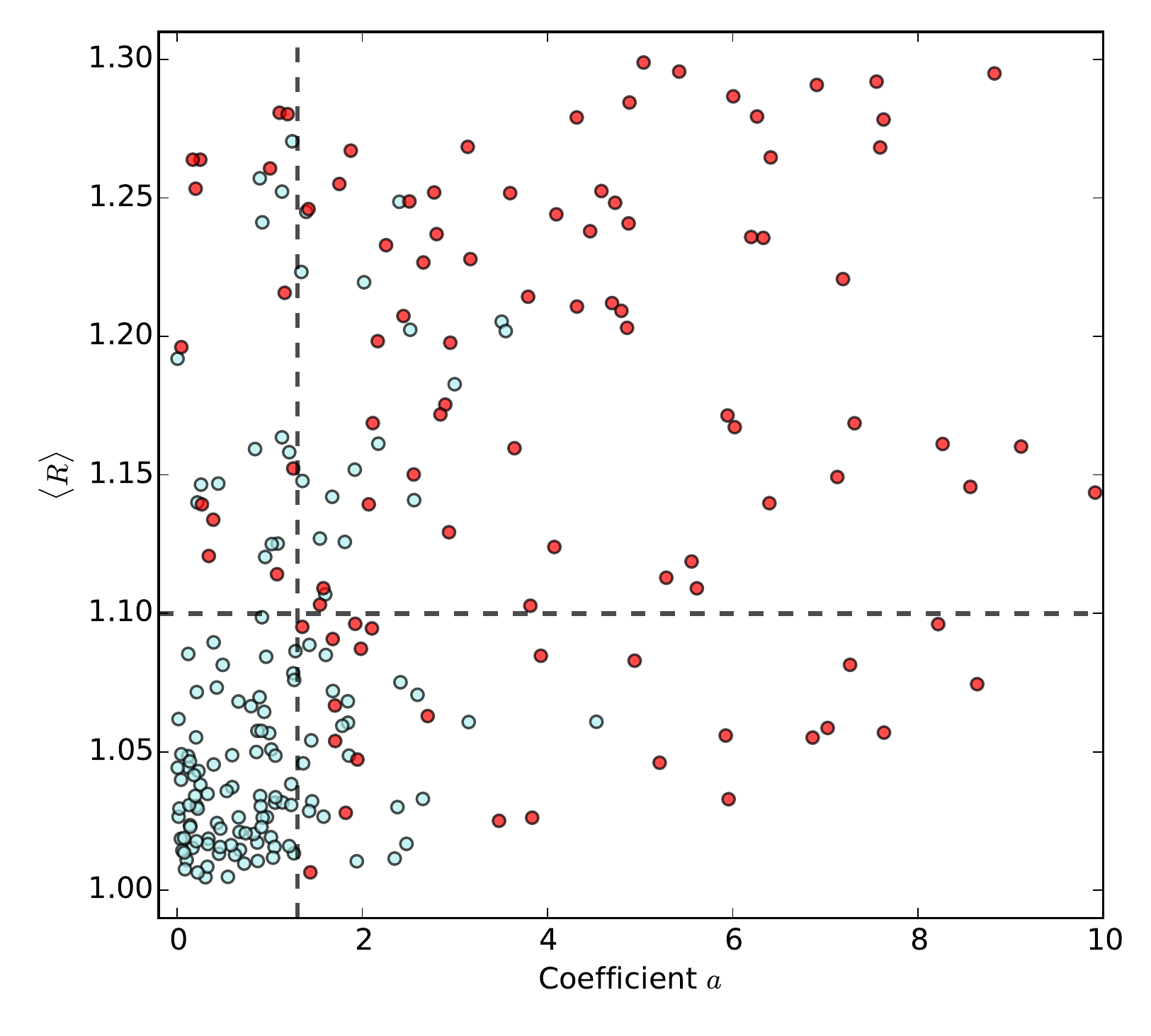}
\caption{Mean spline/quadratic function ratio $\langle R \rangle$ versus the coefficient $a$ of the quadratic function, for all UKIDSS sources satisfying the SED matching criteria of Equation~\ref{eq:sed-match-criteria} and with $a>0$. The sources are color-coded according to their visual evaluation of good (pale sky-blue) or bad (red) match with the MIR SED. The dashed lines indicate the limits on $\langle R \rangle$ and $a$ defining the conditions of Equation~\ref{eq:convex-criteria} to distinguish the unambiguous sources from the ambiguous ones.} 
\label{fig:convex-criteria}
\end{figure}

We could have simply reassigned all sources satisfying the conditions of Equation~\ref{eq:sed-match-criteria} and with $a >0$ to the set of sources not having a good SED match. However, there are some UKIDSS sources matching the MIR SED which are reddened and still produce a slightly convex quadratic fit (as the example in the \emph{right} panel of Figure~\ref{fig:example-convex-sed}), probably due to uncertainties or variability on the measured fluxes. Since the number of sources satisfying the SED matching criteria and with convex quadratic fits is relatively low (246 sources out of a total of 4958 UKIDSS sources matching the MIR SED), the SEDs of these source were visually inspected and evaluated as good or bad matches, similarly to what was done for the validation sample (Section~\ref{sec:sed-match}). In Figure~\ref{fig:convex-criteria}, we plot the parameter $\langle R \rangle$ against the quadratic coefficient $a$ for this sample of 246 sources, color-coded according to their evaluation after the visual inspection. We found that sources within the region defined by the conditions
\begin{equation}
\label{eq:convex-criteria}
\begin{cases} a &\le \,1.3 \\
              \langle R \rangle &\le \,1.1
\end{cases}
\end{equation}
can reliably be kept as having good SED matches, whereas sources outside that area can have either good or bad SED matches. During the visual inspection process, we also noticed that sources outside the region defined by Equation~\ref{eq:convex-criteria} were harder to evaluate. We then decided to consider all sources with $a >0$ and satisfying the conditions of Equation~\ref{eq:sed-match-criteria} but not those of Equation~\ref{eq:convex-criteria} as ambiguous sources, which follow the standard decision rule of Equation~\ref{eq:FA-BA separation}.

We found that the convexity of the quadratic fit was useful to identify what kind of sources can be misclassified as bad SED matches by the SED matching criteria, and in this way we could improve the random reassignment of sources in the MC simulations (see Section~\ref{sec:MC-contamination}), done to estimate the uncertainties on the statistics. For simplicity, we will call the sources with $a > 0$ and not satisfying Equation~\ref{eq:convex-criteria} as \emph{extremely convex} sources, regardless if they satisfy the SED matching criteria or not. All the 11 sources from the validation sample with $\Delta \theta > 0.57\arcsec$ and $\langle R \rangle \le 1.3$ that were visually evaluated as good SED matches have $a < 0$ (10 sources), or have $a > 0$ and satisfy the conditions of Equation~\ref{eq:convex-criteria} (1 source). Therefore, the fraction of extremely convex sources within the whole sample of misclassified sources with $\Delta \theta > 0.57\arcsec$ and $\langle R \rangle \le 1.3$ is low, and probably comparable to the fraction $f_{\rm c}$ of those sources within the set satisfying Equation~\ref{eq:sed-match-criteria}. We thus set the fraction of extremely convex sources within the reassigned sources with $\Delta \theta > 0.57\arcsec$ and $\langle R \rangle \le 1.3$ in each MC simulation to a value close to $f_{\rm c}$, following an analogous method (based on the binomial distribution) to the one used to set the number of reassigned sources (see Section~\ref{sec:MC-contamination}).

\section{Details of SED matching statistics}
\label{sec:statistics-details}

\subsection{Monte Carlo simulations of contamination}
\label{sec:MC-contamination}

As described in Section~\ref{sec:sed-match}, the SED matching criteria defined by Equation~\ref{eq:sed-match-criteria} are affected by some contamination, which introduces uncertainties on the classification of \citetalias{Robitaille2008} objects presented in Section~\ref{sec:statistics}. We estimated these uncertainties by running $10^5$ MC simulations which included the presence of contaminants.

The totality of UKIDSS sources analyzed by the SED matching method were initially divided into good or bad SED matches using the conditions of Equation~\ref{eq:sed-match-criteria}. Then, in each MC simulation, we randomly reassigned a certain fraction of sources satisfying Equation~\ref{eq:sed-match-criteria} to the set of sources having a bad SED match; in the same way, a fraction of sources with $\Delta \theta > 0.57\arcsec$ and $\langle R \rangle \le 1.3$ were reassigned to the set of sources having a good SED match. We assumed no contamination for bad SED matches defined by $\langle R \rangle > 1.3$, as we found for the validation sample in Section~\ref{sec:sed-match}.

For simplicity, we estimated the probability that a source is a contaminant by assuming a Bernoulli trial. Hence, the specific number of reassigned sources for each case was drawn from a binomial distribution $B(n,p)$, where the parameters $n$ and $p$ are, respectively, the total number of UKIDSS sources that initially fall in each set, and the corresponding contamination fraction. For the validation sample, we will generically refer to the total number of sources in each set and the number of misclassified sources as $n_0$ and $k_0$, respectively, so that $n_0 = 199$ and $k_0 = 6$ for the set of initially good SED matches defined by the conditions of Equation~\ref{eq:sed-match-criteria}, and $n_0 = 47$ and $k_0 = 11$ for the set of initially bad SED matches defined by $\Delta \theta > 0.57\arcsec$ and $\langle R \rangle \le 1.3$. Given that the validation sample is just a randomly selected subsample, the corresponding number of misclassified sources does not determine the ``true'' underlying contamination fraction of the whole set, but defines a probability distribution from which we can draw the contamination fraction~$p$ used in each MC simulation. If we assume that the sample space of the validation sample in each set consists of subsets of exactly $n_0$ sources, it can be shown  -- using the Bayes' theorem -- that the resulting distribution function for $p$ given $k_0$ is
\begin{equation}
\label{eq:p-dist-function}
 f(p\mid k_0) = \frac{(n_0 + 1)!}{k_0! (n_0 - k_0)!} p^{k_0} (1 - p)^{n_0 - k_0}~,
\end{equation}
which is equivalent to the beta distribution ${\rm Beta}(\alpha, \beta)$ with parameters $\alpha = k_0 + 1$ and $\beta = n_0 - k_0 + 1$.

In summary, each MC simulation reassigned a certain number $k$ of randomly chosen\footnote{For consistency, the sources from the validation sample that were visually considered as correctly classified, and those with a convex quadratic fit but satisfying the criteria of  reliable sources (see Appendix~\ref{sec:convex}), were excluded from the reassignment, whereas the sources from the validation sample found to be misclassified were always reassigned; however, given the relatively low number of sources in these samples, this probably has, if any, a minor effect on the statistics. In addition, most of the reassigned sources from the set of initially bad SED matches were forced to have concave quadratic fits, to resemble the corresponding misclassified sources from the validation sample (see Appendix~\ref{sec:convex} for details).} sources from each set, where $k$ had been drawn from the binomial distribution $B(n,p)$, and $p$ had been previously drawn from the distribution of Equation~\ref{eq:p-dist-function}. The \citetalias{Robitaille2008} objects were then classified into the categories defined in Section~\ref{sec:statistics}, using the resulting separation of the UKIDSS sources into good or bad SED matches in each MC simulation.

\subsection{Sampling error on the fraction of YSOs and AGB stars}
\label{sec:hypergeometric}

In Section~\ref{sec:statistics}, we also divided the sample into YSOs and AGB stars using the criteria by \citetalias{Robitaille2008}, in order to identify whether one of both populations is more important within a certain category with respect to the others in the classification defined in Table~\ref{tab:classification}. The are two sources of uncertainty on the fraction of YSOs or AGB stars within each category. The first one is the dispersion derived from the MC simulations, produced by the variation of both the total number of objects in a category and its corresponding number of YSOs or AGB stars. The second error arises from the fact that each category is a subsample of the whole set, and therefore the proportions of YSOs and AGB stars can statistically be slightly different than the original ones, even if the category is unbiased regarding the YSO/AGB star separation (i.e., equivalent to a random subsample). The total sample consists of $M= 5355$~\citetalias{Robitaille2008} objects which are divided in $J_{\rm YSOs} = 3825$~YSOs and $J_{\rm AGB} = 1530$~AGB stars; then, any random subsample of $m$~objects will contain $j_{\rm YSOs}$~YSOs and $j_{\rm AGB}$~AGB stars with expected values of $j_i = m J_i/M$, where $i = \{{\rm YSOs, AGB}\}$, and a certain dispersion. This is a direct application of the hypergeometric probability distribution for the random variable $j_i$. The variance of the fraction $j_i/m$ is then given by
\begin{equation}
\label{eq:hypergeometric-variance}
\sigma^2(j_i/m) = \frac{J_i(M - J_i)(M - m)}{m M^2 (M-1)}~.
\end{equation}

The total uncertainty on the fraction of YSOs or AGB stars for a given category was computed adding in quadrature the dispersion from the MC simulations and $\sigma(j_i/m)$ from Equation~\ref{eq:hypergeometric-variance}, where we used the mean number of objects $m = \langle N \rangle$ of the category (see Table~\ref{tab:classification}).

\section{Special and ambiguous cases of SED match: extended categories}
\label{sec:extended-sample}

\begin{table*}[!t]
\renewcommand{\arraystretch}{1.1}
\caption{Statistics for special cases.}
\label{tab:classification-extended}
\centering
\begin{tabular}{lcccccc}
\hline\hline
Category & $\langle N \rangle$ & $\sigma(N)$ & $\%_{\rm YSOs}$ & 
$\sigma(\%_{\rm YSOs})$ & $\%_{\rm AGB}$ & $\sigma(\%_{\rm AGB})$\\
%(1) & (2) & (3) & (4) & (5) & (6) & (7)\\
\hline
object with \verb|BA| source & 898 & 16 & 76.7 & 1.4 & 23.3 & 1.4 \\
\verb|KO10| with linear match & 28 & 1 & 56.1 & 8.6 & 43.9 & 8.6 \\
\verb|KO10| with no linear match & 13 & 2 & 64 & 14 & 36 & 14 \\
\hline
\end{tabular}
\tablefoot{Columns have the same meaning as the ones of Table~\ref{tab:classification}.}
\end{table*}

We can identify some special lower-quality cases of SED match/mismatch, different from the main classification defined in Section~\ref{sec:statistics}, that might still be useful for the statistics of UKIDSS-single objects presented in Section~\ref{sec:dominant-sources}. In Table~\ref{tab:classification-extended}, we distinguish three situations, based on the decision rules detailed in Appendix~\ref{sec:decision-rules}, that were previously grouped in the single category ``ambiguous/special cases'' of Table~\ref{tab:classification}. The first row of Table~\ref{tab:classification-extended} counts all objects that have a bright ambiguous UKIDSS source (\verb|BA| source), and were thus excluded from the SED matching statistics. Objects denoted here as \verb|KO10| represent cases of \verb|KO-2| or \verb|KO-U| sources that are the brightest at $K$ by a factor $>10$ and were evaluated by the secondary SED matching criteria explained in Appendix~\ref{sec:decision-rules}, referred as ``linear match''. \citetalias{Robitaille2008} objects with \verb|KO10| sources with linear match can be therefore considered as UKIDSS-single. Columns of Table~\ref{tab:classification-extended} have the same meaning as the ones of Table~\ref{tab:classification}. We note that the relatively more clustered environment of YSOs could explain the higher fraction of candidate YSOs in objects affected by bad photometry (objects with a \verb|BA| source).

We constructed an extended sample to compute a new estimate of the fraction of UKIDSS-single objects with slightly improved statistics. In addition to the separation into UKIDSS-single and non-UKIDSS single objects explained in Section~\ref{sec:dominant-sources}, here we included the classes \verb|UM_SM_K10|\footnote{In this case, we should exclude the \texttt{UM\_SM\_K10} objects that eventually had a low $S/N$, or had a \texttt{BA} source; however, in our sample this did not occur.} and \verb|KO10| with linear match as possible UKIDSS-single objects, and correspondingly the category \verb|KO10| with no linear match as possible non-UKIDSS-single. These categories are very rare, but we also considered here the \citetalias{Robitaille2008} objects affected by peripheral saturation (see Section~\ref{sec:quality}), since the PSF wings of nearby saturated stars should not severely affect the photometry of UKIDSS sources with $S/N > 30$ at $K$, which was required for our SED matching statistics. This extended sample comprises a total of $\sim 1440$~AGB star candidates and $\sim 2950$~YSO candidates, representing an increase of over 10\% in the sample size. The fraction of UKIDSS-single objects are $92.0 \pm 1.2 \%$ for candidate AGB stars, and $87.2 \pm 1.5 \%$ for candidate YSOs, which are identical within the uncertainties to the fractions estimated for the good-quality sample of Section~\ref{sec:dominant-sources}.

\end{appendix}

\end{document}